\documentclass[a4paper, 11pt]{article}
\RequirePackage[OT1]{fontenc}
\usepackage{amsfonts,amssymb,graphics,epsfig,verbatim,bm,latexsym,amsmath,url,amsbsy,multirow,rotating,tikz,enumerate,subfig}
\usepackage[textheight=722pt,textwidth=500pt,centering,headheight=50pt,headsep=12pt,footskip=12pt]{geometry}
\usepackage[colorlinks,citecolor=blue,linkcolor=purple!60!black,urlcolor=purple!60!black]{hyperref}
\usepackage[framemethod=TikZ]{mdframed}
\usepackage{mathtools}
\usetikzlibrary[shapes.misc]
\usetikzlibrary{arrows,automata}
\allowdisplaybreaks

\begin{document}
\begin{center}
\textbf{{\LARGE{Using Longitudinal Targeted Maximum Likelihood Estimation in Complex Settings with Dynamic Interventions}}}
\end{center}\vspace*{0.5cm}

{\Large
\begin{center}
Michael Schomaker\footnote{Centre for Infectious Disease Epidemiology \& Research, University of Cape Town, Cape Town, South Africa, \href{mailto:michael.schomaker@uct.ac.za}{michael.schomaker@uct.ac.za} and\\ \hspace*{0.4cm} Institute of Public Health, Medical Decision Making and Health Technology Assessment, UMIT - University for Health Sciences, Medical Informatics and Technology, Hall in Tirol, Austria, \href{mailto:michael.schomaker@umit.at}{michael.schomaker@umit.at}},
Miguel Angel Luque-Fernandez\footnote{Biomedical Research Institute of Granada -- Noncommunicable and Cancer Epidemiology Group, Andalusian School of Public Health, University of Granada, Granada, Spain, \href{mailto:miguel.luque.easp@juntadeandalucia.es}{miguel.luque.easp@juntadeandalucia.es}
%and \\ \hspace*{0.4cm} Department of Epidemiology and Population Health, London School of %Hygiene and Tropical Medicine, London, United Kingdom, %\href{Miguel-Angel.Luque@lshtm.ac.uk}{Miguel-Angel.Luque@lshtm.ac.uk} and \\ \hspace*{0.4cm} %Department of Epidemiology, Harvard T.H Chan School of Public Health, Boston, United States, %\href{mailto:mluquefe@hsph.harvard.edu}{mluquefe@hsph.harvard.edu}
 }
Valeriane Leroy\footnote{INSERM, Unite 1027, Toulouse, France, \href{mailto:Valeriane.Leroy@isped.u-bordeaux2.fr}{Valeriane.Leroy@isped.u-bordeaux2.fr}},
Mary-Ann Davies\footnote{Centre for Infectious Disease Epidemiology \& Research, University of Cape Town, Cape Town, South Africa, \href{mailto:mary-ann.davies@uct.ac.za}{mary-ann.davies@uct.ac.za}}
\end{center}
}

\begin{abstract}
Longitudinal targeted maximum likelihood estimation (LTMLE) has very rarely been used to estimate dynamic treatment effects in the context of time-dependent confounding affected by prior treatment when faced with long follow-up times, multiple time-varying confounders, and complex associational relationships simultaneously. Reasons for this include the potential computational burden, technical challenges, restricted modeling options for long follow-up times, and limited practical guidance in the literature. However, LTMLE has desirable asymptotic properties, i.e. it is doubly robust, and can yield valid inference when used in conjunction with machine learning. It also has the advantage of easy-to-calculate analytic standard errors in contrast to the g-formula, which requires bootstrapping. We use a topical and sophisticated question from HIV treatment research to show that LTMLE can be used successfully in complex realistic settings and we compare results to competing estimators. Our example illustrates the following practical challenges common to many epidemiological studies 1) long follow-up time (30 months), 2) gradually declining sample size 3) limited support for some intervention rules of interest 4) a high-dimensional set of potential adjustment variables, increasing both the need and the challenge of integrating appropriate machine learning methods 5) consideration of collider bias. Our analyses, as well as simulations, shed new light on the application of LTMLE in complex and realistic settings: we show that (i) LTMLE can yield stable and good estimates, even when confronted with small samples and limited modeling options; (ii) machine learning utilized with a small set of simple learners (if more complex ones cannot be fitted) can outperform a single, complex model, which is tailored to incorporate prior clinical knowledge; (iii) performance can vary considerably depending on interventions and their support in the data, and therefore critical quality checks should accompany every LTMLE analysis. We provide guidance for the practical application of LTMLE.
\end{abstract}

\begin{mdframed}[backgroundcolor=red!10, linecolor=black!50]
{\small
The published version of this working paper can be cited as follows:\\[0.25cm]
Schomaker, M., Luque-Fernandez, M., Leroy, V., Davies, M.A.\\
\textit{Using Longitudinal Targeted Maximum Likelihood Estimation in Complex Settings with Dynamic Interventions}\\
Statistics in Medicine (2019); 38:4888-4911\\
\url{https://doi.org/10.1002/sim.8340}
}
\end{mdframed}

\clearpage
\section{Introduction}
Causal quantities, such as the average treatment effect, causal odds ratio or causal risk ratio, can often be estimated from observational data if assumptions about the data generating process are made and met. In this context, Targeted Maximum Likelihood Estimation (TMLE) is one estimation option. Since the mid 2000's TMLE has been implemented successfully in numerous applications, initially for the effects of treatments or exposures at a single time point \cite{Gruber:2010b, Pearl:2016, Luque:2017, Luque:2018, Schomaker:2018}. In recent years, an increasing body of work has dealt with the application of TMLE in longitudinal settings \cite{Decker:2014, Gianfrancesco:2016, Hubbard:2012, Kreif:2017, Lendle:2017, Neugebauer:2014, Petersen:2014, Petersen:2014a, Schnitzer:2014, Schnitzer:2014b, Schnitzer:2016, Stitelman:2012, Tran:2016, Tran:2019}.

Longitudinal targeted maximum likelihood estimation (LTMLE) requires the fitting of models for the outcome, censoring, and treatment mechanisms, iteratively at each point in time. Alternative approaches include those based on estimating the treatment and censoring mechanism only (inverse probability of treatment weighted approaches)]\cite{Robins:2000}, the iterated outcome regressions only (one form of the g-formula)\cite{Bang:2005} or an outcome regression combined with the conditional densities of the confounders (``traditional'' parametric g-formula)\cite{Robins:1986}. LTMLE is a doubly robust method which means that the causal quantity of interest is estimated consistently if either the iterated outcome regressions or the treatment and censoring mechanism are estimated consistently \cite{vanderLaan:2012}. Moreover, LTMLE allows analytic estimation of standard errors and confidence intervals, based on estimates of the influence curve, as opposed to the g-formula which makes use of bootstrapping \cite{vanderLaan:2011, Stefanski:2002}. To increase the likelihood of correct model specification, the use of machine learning algorithms are typically recommended. LTMLE, like other doubly robust methods, has the advantage over both inverse probability weighted and g-formula methods that it can more readily incorporate machine learning methods while retaining valid statistical inference.

The current literature is somewhat limited with respect to understanding the implications of using LTMLE for long follow-up times when dealing with small (or moderate) sample sizes and complex relationships among a multitude of confounders. Most studies to date were illustrative in nature and have dealt with a small number of follow-up time points, i.e. two to nine \cite{Decker:2014, Gianfrancesco:2016, Hubbard:2012, Kreif:2017, Lendle:2017, Petersen:2014, Petersen:2014a, Schnitzer:2014, Schnitzer:2014b, Tran:2016, Tran:2019}, or large sample sizes \cite{Neugebauer:2014, Schnitzer:2014b, Tran:2016, Tran:2019}, but to our knowledge very rarely with small (or moderate) sample sizes and long follow-up. A notable exception is the study from Schnitzer et al \cite{Schnitzer:2016} with a very long follow-up period, but the authors reported memory problems and had difficulties dealing with a decreasing sample size and sparsity of events.  In spite of LTMLE's double robustness feature, consistent estimation of both treatment and censoring mechanisms and outcome regressions, with reasonable rates of convergence, is important to improve efficiency, reduce finite sample bias, and provide the basis for valid statistical inference. A small sample may limit the application of data-adaptive approaches, such as additive models, boosting, random forests, shrinkage estimators, support vector machines and others, because the selection of tuning parameters and convergence of fitting algorithms may not work well, or not at all, under these circumstances -- particularly when confronted with multiple time-dependent confounders which influence outcome and treatment non-linearly and in interaction with each other. Interestingly, in many studies using LTMLE the application of machine learning algorithms is limited  \cite{Gianfrancesco:2016, Hubbard:2012, Neugebauer:2014, Stitelman:2012, Tran:2016}, posing the question whether this was done intentionally because of restricted sample size or computational limitations.

A common estimation approach for LTMLE requires model fitting on the subset of subjects who follow the intervention of interest (and remain alive and uncensored). This may pose a problem if follow-up time is long and therefore the sample size gradually reduces: under these circumstances it may not longer be possible to successfully fit a multitude of learners which makes correct model specification more challenging. While there are alternative estimation approaches \cite{Petersen:2014} which utilize the full sample, it remains to be investigated how the performance of these estimators differ in a finite sample.

For applying LTMLE in survival analyses, both the treatment and censoring mechanisms need to be modeled; in fact, for $T$ time points, the cumulative inverse probabilities of remaining uncensored and following the intervention of interest, i.e. the product of $T\times2$ inverse probabilities, are part of the weighted fitting procedure. Unstable weights, possibly dealt with by truncation of the probabilities on which the weight calculation is based, could therefore emerge under long follow-up and non-ideal modeling strategies. It remains unclear to what degree this is a threat to the successful implementation of LTMLE.

The present article is based on our motivating question outlined in the next section. To answer it, one has to deal with long follow-up times, gradually reduced sample size (down to about 10\% of the original sample size of 2604) and multiple time-dependent confounders. Clinical knowledge suggests that the data-generating process is complex in the sense that modeling should go beyond standard regression approaches. We investigate challenges and solutions with appying LTMLE in this setting, both by means of simulations and data analyses. In particular, we try to answer the following questions: (i) is it feasible to apply LTMLE in such a setting; (ii) if yes, will the complex setup reduce the applicability of super learning and would this impact the success of LTMLE?; (iii) under such a challenging scenario, could a manual implementation which is informed by prior clinical knowledge and is not data-adaptive be more successful than an automated LTMLE procedure?; (iv) how do long follow-up times, gradually reduced sample size, and truncated inverse probabilities impact on LTMLE for different intervention mechanisms? (v) what can be learned from these points that would benefit future analyses which attempt to apply LTMLE in complex data settings?

We introduce the motivating question in Section \ref{sec:motivation}, followed by the theoretical framework in Section \ref{sec:framework}. After presenting our data analysis (Section \ref{sec:data_analysis}) we discuss the implications of LTMLE in such a setting by means of both practical considerations and simulated studies in the following two sections. We conclude in Section \ref{sec:conclusion}.

\iffalse
\cite{Decker:2014} t=2 ; n=530; ?tdc \\
\cite{Gianfrancesco:2016} t=9 ; n=952; limited super learner; 7 tdc\\
\cite{Hubbard:2012} t=2; n=165; limited super learner; 2tdc\\
\cite{Kreif:2017} t=7 ; n=1369; good super learner; many tdc\\
\cite{Neugebauer:2014} t=36; n=51179; limited super learner \\
\cite{Schnitzer:2014} t=6; n=780 \\
\cite{Schnitzer:2014b} t=6; n=17000\\
\cite{Stitelman:2012} t=15+?; n=650; 54 LTFU; no super learning(?) - simplified model assumptions (?); 2 tdc\\
\cite{Tran:2016} t=4; n=16479; reported to prefer not too many time-points, thus pragmatic intervals chosen; many tdc \\
\cite{Petersen:2014} t=2; n=2411
\fi

\section{Motivating Example}\label{sec:motivation}
During the last decade the World Health Organization (WHO) updated their recommendations regarding the use of antiretroviral drugs for treating and preventing HIV infection several times. In the past, antiretroviral treatment (ART) was only given to a child if his/her measurements of CD4 lymphocytes fell below a critical value or if a clinically severe event (such as tuberculosis or persistent diarrhoea) occurred. Based on both increased knowledge from trials and causal modeling studies, as well as pragmatic and programmatic considerations, these criteria have been gradually expanded to allow earlier treatment initiation in children: ART has shown to be effective and to reduce mortality in infants and adults \cite{Edmonds:2011, Violari:2008}, but concerns remain due to a potentially increased risk of toxicities, early development of drug resistance, and limited future options for people who fail treatment. In late 2015, WHO decided to recommend immediate treatment initiation in all children and adults.

It remains important to investigate the effect of different treatment initiation rules on mortality, morbidity and child development outcomes; for example, because of possible unknown long-term treatment effects and the increase in drug related toxicities. However given the shift in ART guidelines towards earlier treatment initiation it is no longer ethically possible to conduct a trial to answer this question. Moreover, trials are often limited by testing interventions in idealized, non-realistic, settings -- for example by imposing regular medical visits and therefore additional personal attention which would not happen in most public health programmes.

Thus, observational data can be used to obtain effect estimates to compare early versus delayed treatment initiation rules. Methods such as inverse probability weighting of marginal structural models, the g-formula, and longitudinal targeted maximum likelihood estimation can be used to obtain results in complicated longitudinal settings where time-varying confounders affected by prior treatment are present --- such as, for example, CD4 count which influences both the probability of ART initiation and outcomes \cite{Daniel:2013, Petersen:2014}.

Below we focus on children's growth outcomes, measured by height-for-age z-scores (HAZ), and explore the effect of different treatment initiation criteria on HAZ. The PREDICT trial, conducted among Asian children aged 1-12, suggested a growth difference between early and deferred treatment initiation, among those children who survived \cite[Table 3]{Phutanakit:2012}. However, unfortunately, the study was underpowered and had other limitations, including few very young children. In our analysis, we are particularly interested in whether strategies of earlier treatment initiation are beneficial for children's growth.

\section{Methodological Framework}\label{sec:framework}

Consider $n$ subjects studied at baseline ($t=0$) and during discrete follow-up times ($t=1,\ldots,T$) where $T$ defines the end of follow-up. Some subjects may be censored and not be followed until $T$. The data for each time point $t$ consists of the outcome $Y_t$, an intervention variable $A_t$, $q$ time-dependent covariates $\mathbf{L}_t=\{L^1_t,\ldots,L^q_t\}$, and a censoring indicator $C_t$. Note that $\mathbf{L}_t$ contains indicator variables which indicate whether each covariate is measured for the individual at time $t$ or not. Baseline variables, i.e. variables measured at $t=0$, are denoted as $\mathbf{L_0}=\{L^1_0,\ldots,L^{q_0}_0\}$. Treatment and covariate histories of an individual $i$ up to and including time $t$ are represented as $\bar{A}_{t,i}=(A_{0,i},\ldots,A_{t,i})$ and $\bar{L}^s_{t,i}=(L^s_{0,i},\ldots,L^s_{t,i})$ respectively. $C_t$ equals $0$ if a subject gets censored in the interval $(t-1,t]$, and $1$ otherwise. Therefore, $\bar{C}_t=1$ is the event that an individual remains uncensored until time $t$. Further, let $S_{t,i}$ be an indicator variable which is $1$ if a patient $i$ is still alive at time $t$ and $0$ otherwise.

The counterfactual outcome $Y_{t,i}^{\bar{a}_{t}}$ refers to the hypothetical outcome that would have been observed at time $t$ if subject $i$ had received, possibly contrary to the fact, the treatment history $\bar{A}_{t,i}=\bar{a}_t$. Similarly, $\mathbf{L}_{t,i}^{\bar{a}_{t}}$ are the counterfactual covariates related to the intervention $\bar{A}_{t,i}=\bar{a}_t$. The above notation refers to \textit{static} treatment rules; a treatment rule may, however, depend on covariates, and in this case it is called \textit{dynamic}. A dynamic rule ${d}_t(\mathbf{\bar{L}}_{t})=\bar{d}_t$ assigns treatment ${A}_{t,i} \in \{0,1\}$ as a function of the time-dependent confounder history $\mathbf{\bar{L}}_{t,i}$; and can also assign survival $S_{t,i}=1$ and non-censoring $C_{t,i}=1$ if required. The notation $\bar{A}_t = \bar{d}_t$ refers to the treatment (and censoring) history up to and including time $t$ according to the rule $\bar{d}_t$.  The counterfactual outcome at time $t$ related to a dynamic rule $\bar{d}_t$ is $Y_{t,i}^{\bar{d}_t}$, and the counterfactual covariates at the respective time point are $\mathbf{L}_{t,i}^{\bar{d}_t}$.

\subsection{Target quantity}
We choose $\psi_T = \mathbb{E}(Y_T^{\bar{d}_t})$ as the target quantity. More specifically, we are interested in the expected value of $Y$ at time $T$ under an intervention rule $\bar{d}_t$ which assigns $A_t$ as a function of $\bar{\mathbf{L}}_t$ and sets $C_t$ and $S_t$ deterministically to $1$; that is, we consider rules which intervene on $\mathbf{A}_t=\{A_t,C_t,S_t\}$. The choice of this target quantity is based on the motivating data example and justified in more detail in Section \ref{sec:data_analysis}.

Below, in Sections \ref{sec:gformula}, \ref{sec:sgformula}, and \ref{sec:tmle}, we are going to introduce 3 different estimators which are generally suitable to estimate $\psi_T$. These estimators are further discussed in Section \ref{sec:presentation}, and are implemented and compared in both the data analysis (Section \ref{sec:data_analysis}) and the simulation study (Section \ref{sec:simulation}).

\subsection{The g-formula}\label{sec:gformula}
The quantity of interest is identifiable through the g-formula. \cite{Robins:1986},

\begin{eqnarray}\label{eqn:gformula}
&\psi_T=&\mathbb{E}(Y_T^{\bar{d}})
=\int_{\mathbf{\bar{l}}\in\mathbf{\bar{L}}_t}
\left\{
\begin{aligned}
&\mathbb{E}(Y_T|\bar{\mathbf{A}}_{T-1}=\bar{d}_{T-1}, \mathbf{\bar{L}}_T=\mathbf{\bar{l}}_T
)\times\\
&\prod_{t=1}^T
  \begin{array}{l}
  f(\mathbf{l}_t|\bar{\mathbf{A}}_{t-1}=\bar{d}_{t-1}, \mathbf{\bar{L}}_{t-1}=\mathbf{\bar{l}}_{t-1})
  \end{array}
\end{aligned}
\right\}
\, d\mathbf{\bar{l}} \,,
\end{eqnarray}

where $f(\cdot)$ refers to the probability density function of the respective random variable(s), $Y_t \in \mathbf{L}_t$ ($t<T$) and $\mathbf{A}_t=\{A_t,C_t,S_t\}$. It is an option to assume an order for the variables $\mathbf{L}_t=\{L^1_t,\ldots,L^q_t\}$ in which case we could write the joint distribution $f(\mathbf{l}_t|\bar{\mathbf{A}}_{t-1}=\bar{d}_{t-1}, \mathbf{\bar{L}}_{t-1}=\mathbf{\bar{l}}_{t-1})$ in (\ref{eqn:gformula}) as
\begin{eqnarray}\label{eqn:gformula2}
\prod_{s=1}^{q}
f(l_t^s|\bar{\mathbf{A}}_{t-1} = \bar{d}_{t-1}, \mathbf{\bar{L}}_{t-1}=\mathbf{\bar{l}}_{t-1}, L_t^1 = l_t^1,\ldots, L_t^{s-1}=l_t^{s-1})\,.
\end{eqnarray}

The factorization in (\ref{eqn:gformula2}) makes an implicit assumption about the time ordering of the relevant variables. For example, if we have 3 time-dependent confounders ($q=3$), we assume that the order is $L^1_t \rightarrow L^2_t \rightarrow L^3_t$. More generally we assume $L^1_t \rightarrow L^2_t \rightarrow L^3_t \rightarrow Y_t \rightarrow A_t \rightarrow C_t \rightarrow S_t$ for $t=0,1,\ldots,T$, see Sections \ref{sec:data_analysis_setting} and \ref{sec:MvM_structure} for more details on the chosen time ordering. We further assume time points to represent intervals and thus both covariates and a fatal event like death can potentially be measured at the same time t.

The equality holds under several assumptions. One is \textit{consistency} which says that the counterfactuals equal the observed data under assignment to the treatment actually taken; more technically, $Y^{\bar{d}_t}_T = Y_T$ if $\bar{\mathbf{A}}_{t-1} = \bar{d}_{t-1}$ and $\bar{\mathbf{L}}_t^{\bar{d}_t}=\bar{\mathbf{L}}_t$ if $\bar{\mathbf{A}}_{t-1} = \bar{d}_{t-1}$. Another assumption is (sequential) conditional exchangeability (or \textit{no unmeasured confounding}) which requires the counterfactual outcome under the assigned treatment rule to be independent of the observed treatment assignment, given the observed past: $Y^{\bar{d}_t}_T\coprod {\mathbf{A}_{t-1}|\bar{\mathbf{L}}_t, \bar{\mathbf{A}}_{t-2}}$ for $\forall \bar{\mathbf{A}}_t=\bar{d}_t, \bar{\mathbf{L}}_t=\bar{\mathbf{l}}_t, t = 0,\ldots,T$. \textit{Positivity} is the requirement that individuals have a positive probability of continuing to receive treatment according to the assigned treatment rule, given that they have done so thus far and irrespective of the covariate history: $P(\mathbf{A}_t=\bar{d}_t|\bar{\mathbf{L}}_t=\bar{\mathbf{l}}_t,\bar{\mathbf{A}}_{t-1}=\bar{d}_{t-1})>0$ for $\forall t,\bar{d}_t,\bar{\mathbf{l}}_t$ with $P(\bar{\mathbf{L}}_t=\bar{l}_t,\bar{\mathbf{A}}_{t-1}=\bar{d}_{t-1}) \neq 0$.

More comprehensive interpretations of the above assumptions can be found in \cite{Daniel:2013, Robins:2009, Daniel:2011, Young:2011, Robins:2004} and are discussed in Section \ref{sec:MvM} as well.

\subsubsection{Practical implementation of the (parametric) g-formula:}\label{sec:gformula_implementation}

the integral (\ref{eqn:gformula}) can be approximated by means of simulation, as shown previously \cite{Young:2011, Schomaker:2017}:

\begin{enumerate}
\item Estimate the conditional densities in (\ref{eqn:gformula}) by means of suitable, possibly semi-parametric, regression models. For example, if there are 3 time-dependent covariates one needs to fit 3 models for each time point, as well as a model for $Y_t$ for each point in time (because $Y_t \in \mathbf{L}_t \,\text{for}\, t<T$ and for $t=T$ one needs to estimate $\mathbb{E}(Y_T|\bar{\mathbf{A}}_{T-1}=\bar{d}_{T-1}, \mathbf{\bar{L}}_T=\mathbf{\bar{l}}_T$). Note that for this step some parametric assumptions are required, e.g. committing to a normal distribution of the residuals with homoscedastic variance when using a linear (additive) model.
\item Choose a particular intervention rule $\bar{d}^j_t$.
\item Simulate data \textit{forward in time} for the chosen intervention rule, based on the estimated conditional distributions from Step 1:
     \begin{enumerate}[i)]
      \item for $t=0$, the data are the observed data.
      \item for $t>0$:
       \begin{enumerate}[a)]
       \item set $\mathbf{A}_{t-1}$ according to the treatment rule $\bar{d}^j_t$ (we write $\mathbf{A}_t = \mathbf{A}^{\bar{d}^j_t}_{t-1}$).
       \item take the fitted models from step 1 to estimate the densities $f(l_t^s|\bar{\mathbf{A}}_{t-1} = \bar{d}_{t-1}^j, \mathbf{\bar{L}}_{t-1}=\mathbf{\bar{l}}_{t-1}, L_t^1 = l_t^1,\ldots, L_t^{s-1}=l_t^{s-1})$ under $\bar{d}^j_t$. For instance, when using the fit of a linear model $\mathcal{M}$ where the outcome $Y_{t}$ is regressed on its past, set $\mathbf{A}_{t-1} = \mathbf{A}^{\bar{d}^j_t}_{t-1}$, use $\mathbf{L}_0$ from the observed data and $\mathbf{L}_t$ from the counterfactual estimates $\hat{\mathbf{L}}^{\bar{d}^j_t}_t$ obtained in c) -- and calculate $\mu_{\mathcal{M}}$ and $\sigma_{\mathcal{M}}$ under this intervention and these data based on the fitted model parameters.
       \item then, generate stochastic draws of the conditional distributions under the respective intervention. For example, for a linear regression model as explained in b), draw from a $N(\mu_{\mathcal{M}},\sigma_{\mathcal{M}})$ distribution. This draw relates to the estimated counterfactual $\hat{Y}_{t}^{\bar{d}^j_t}$. The counterfactual estimates  $\hat{\mathbf{L}}^{\bar{d}^j_t}_t$ can be obtained in the same way.
       \end{enumerate}
      \end{enumerate}
\item Steps 1 to 3 produce a counterfactual data set under $\bar{d}^j_t: \mathcal{D}=\{\mathbf{A}^{\bar{d}^j_t}, \mathbf{\hat{L}}^{\bar{d}^j_t}, \hat{Y}^{\bar{d}^j_t}\}$. Thus, $\hat{\psi}_T = \frac{1}{n} \sum_{i=1}^n \hat{Y}^{\bar{d}^j_t}_{T,i}$ can be easily calculated from this data set.
\item Confidence intervals are then created by means of bootstrapping.
\end{enumerate}

\subsection{The sequential g-formula}\label{sec:sgformula}
The g-formula integrates out time-dependent confounders, which are affected by prior treatment, with respect to their post intervention distribution. This can be interpreted as a standardization with respect to these confounders. The sequential g-formula, introduced below, shares the idea of standardization in the sense that one sequentially marginalizes the distribution with respect to $\mathbf{L}$ given the intervention rule of interest \cite{Bang:2005, vanderLaan:2012}.  It is just a re-expression of the traditional g-formula where integration with respect to $\mathbf{L}$ is not needed \cite{Kreif:2017, Petersen:2014a}. The sequential g-formula is also known as the sequential g-computation estimator \cite{vanderLaan:2018} or the iterated conditional expectation estimator \cite{Tran:2019}. In general, using the iterative conditional expectation rule (together with the above stated assumptions of sequential conditional exchangeability, consistency, positivity, time ordering $L^1_t \rightarrow L^2_t \rightarrow L^3_t \rightarrow Y_t \rightarrow A_t \rightarrow C_t \rightarrow S_t$, $t=1,\ldots,T$, $Y_t \in \mathbf{L}_t \,\text{for}\, t<T$ and $\mathbf{A}_t=\{A_t,C_t,S_t\}$) it holds that
\begin{eqnarray}\label{eqn:seq_g_formula}
\mathbb{E}(Y_T^{\bar{d}}) &=& \mathbb{E}(\,\mathbb{E}(\,\ldots\mathbb{E}(\,\mathbb{E}(Y_T|\bar{\mathbf{A}}_{T-1}=\bar{d}_{T-1}, \mathbf{\bar{L}}_T) | \bar{\mathbf{A}}_{T-2}=\bar{d}_{T-2}, \mathbf{\bar{L}}_{T-1}\,)\ldots|\bar{A}_{0}={d}_{0}, \mathbf{{L}}_{1}\,)|\mathbf{{L}}_{0}\,)\,)\,.
\end{eqnarray}

\subsubsection{Practical implementation of the sequential g-formula:}
using equation (\ref{eqn:seq_g_formula}) the target parameter can be estimated as follows:

For $t=T,...,1$:
\begin{enumerate}
\item Use an appropriate model to estimate $\mathbb{E}(Y_t|\mathbf{\bar{A}}_{t-1}, \mathbf{\bar{L}}_t)$. The model is fitted on all subjects that are uncensored and alive (until $t-1$). Note that the outcome refers to the measured outcome for $t=T$ and to the prediction (of the conditional outcome) from step 2 (of iteration $t-1$) if $t<T$.
\item Now, plug in $\bar{\mathbf{A}}_{t-1}=\bar{d}_{t-1}$ based on rule $\bar{d}_t$ and use the regression model from step 1 to predict $Y_t$ at time $t$, which we denote as $\tilde{Y}^{\bar{d}_t}_t$.
\end{enumerate}
Then:
\begin{enumerate}
\setcounter{enumi}{2}
\item The estimate $\hat{\psi}_T$ is obtained by calculating the mean of the predicted outcome from step 2 (where $t=1$).
\item Confidence intervals can be obtained using bootstrapping (or equation (\ref{eqn:ltmle_IC}) below, which however may not yield nominal coverage).
\end{enumerate}

Appendix \ref{sec:appendix_ltmle_man}, which illustrates implementation details for the LTMLE estimator from Section \ref{sec:tmle}, can also be seen as a guideline to implement the sequential g-formula: omitting step 3  from the code yields the desired solution.

\subsection{The LTMLE estimator}\label{sec:tmle}
The LTMLE estimator, which is based on (\ref{eqn:seq_g_formula}), was suggested by van der Laan and Gruber \cite{vanderLaan:2012} who developed the asymptotic theory and algorithm of LTMLE. Their work builds on earlier work  from Bang and Robins \cite{Bang:2005} who suggested a semi-parametric estimator that is defined through parametric regression models. For our target parameter $\psi_T$ the LTMLE framework can be applied as follows: estimate $\psi_T$ with the sequential g-formula, but with an additional \textit{targeted} step for each $t$ which enables doubly robust inference with respect to $\psi_T$, the quantity of interest.

\clearpage
\subsubsection{Practical implementation of the LTMLE estimator:}\label{sec:tmle_implementation} \quad

For $t=T,...,1$:
\begin{enumerate}
\item Use an appropriate regression model to estimate $\mathbb{E}(Y_t|\mathbf{\bar{A}}_{t-1}, \mathbf{\bar{L}}_t)$. The model is fitted on all subjects that are uncensored and alive (until $t-1$). Note that the outcome refers to the measured outcome for $t=T$ and to the prediction (of the conditional outcome) from step 3d (of iteration $t-1$) if $t<T$. (``$Q^0$ model'')
\item Now, plug in $\mathbf{\bar{A}}_{t-1}={\bar{d}}_{t-1}$ based on rule $\bar{d}_t$ and use the regression model from step 1 to predict the outcome at time $t$, i.e. $\tilde{Y}^{\bar{d}_t}_t$.
\item To improve estimation with respect to the target parameter $\psi_t$ update the initial estimate of step 2 by means of the following regression:
\begin{enumerate}[a)]
     \item the outcome refers again to the measured outcome for $t=T$ and to the prediction from item 3d (of iteration $t-1$) if $t<T$.
     \item the offset is the original predicted outcome from step 2 (iteration $t$).
     \item the estimated ``clever covariate'' refers to the cumulative product of inverse treatment and censoring probabilities:
     \begin{eqnarray}\label{eqn:clever_covariate}
     \hat{H}(\bar{A},\bar{C},\mathbf{\bar{L}})_{t-1} = \prod_{s=0}^{t-1} \frac{I(\bar{{A}}_s=\bar{d}_{s}) \times I(\bar{C}_s=1) \times I(\bar{S}_s=1)}{
     \splitfrac{\splitfrac{
     \hat{\mathbb{P}}({A}_s=\bar{d}_{s}| \mathbf{\bar{L}}_{s}=\mathbf{\bar{l}}_{s},\bar{S}_s=1,\bar{C}_{s-1}=1,\bar{{A}}_{s-1}=\bar{d}_{s-1})}{
     \times \hat{\mathbb{P}}({C}_s=1| \bar{{A}}_{s}=\bar{d}_{s}, \mathbf{\bar{L}}_{s}=\mathbf{\bar{l}}_{s},\bar{S}_s=1,\bar{C}_{s-1}=1)
     }}{\times \hat{\mathbb{P}}({S}_s=1| \bar{C}_{s}=1, \bar{{A}}_{s}=\bar{d}_{s}, \mathbf{\bar{L}}_{s}=\mathbf{\bar{l}}_{s},\bar{S}_{s-1}=1)}
     }
     \end{eqnarray}
     Before (\ref{eqn:clever_covariate}) can be calculated the treatment, censoring and survival models in the denominator of (\ref{eqn:clever_covariate}) need to be fitted first. If it is assumed that $S_t$ follows directly after $C_t$, then it is an option to assume and model only one censoring process per time point. We follow this approach in the data analysis for computational convenience (Appendix \ref{sec:appendix_ltmle_man}).
     \item Predict the (updated) outcome, $\tilde{Y}_t^{\bar{d}_t}$, based on the model defined through 3a, 3b, and 3c.
     \end{enumerate}
          This model contains no intercept and is fitted and evaluated for all subjects that are uncensored, alive (until $t-1$), \textit{and follow treatment rule} $\bar{\mathbf{A}}_{t-1}=\bar{d}_{t-1}$ (``$Q^1$ model''). Alternatively, the same model can be fitted with $\hat{H}(\bar{A},\bar{C},\mathbf{\bar{L}})_s$ as a weight rather than a covariate \cite{Kreif:2017, Tran:2019}. In this case an intercept is required. We follow the latter approach in our implementations of Section \ref{sec:data_analysis} and \ref{sec:simulation}.
\end{enumerate}
Then:
\begin{enumerate}
\setcounter{enumi}{3}
\item The estimate $\hat{\psi}_T$ is obtained by calculating the mean of the predicted outcome from step 3d (where $t=1$).
\item Confidence intervals can, for example, be obtained using the vector of the estimated influence curve \cite{vanderLaan:2012} of $\psi_T$, which can be written as
\begin{eqnarray}\label{eqn:ltmle_IC}
\widehat{\text{IC}}(\psi_T) &=& \left\{\sum_{s=1}^T  \hat{H}(\bar{A},\bar{C},\mathbf{\bar{L}})_{s-1}  \left[\tilde{Y}_s^{\bar{d}_t} - \tilde{Y}_{s-1}^{\bar{d}_t} \right] \right\} + \tilde{Y}_0^{\bar{d}_t} - \hat{\psi}_{T,\text{TMLE}}\,,
\end{eqnarray}
where $\tilde{Y}_s$ at time $s=T$ is $Y_T$. An asymptotically normal 95\% confidence interval is then given by
\begin{eqnarray*}
\left[\hat{\psi}_{\text{TMLE}} \pm 1.959964 \sqrt{{\widehat{\text{Var}}(\widehat{\text{IC}})}/{{n}} } \right]  \,.
\end{eqnarray*}
\end{enumerate}

It has been shown that using the five steps above yields a doubly robust estimator of $\psi_T$ \cite{Bang:2005}. Doubly robust means that the estimator is consistent as long as either the outcome model[s] (step 1) or the treatment model[s] (step 3c) are estimated consistently. If both are estimated consistently at reasonable rates, the estimator is asymptotically efficient. This is essentially because step 3, in particular the construction of the covariate in step 3c, guarantees that the estimating equation corresponding to the efficient influence curve is solved, which in turn yields desirable (asymptotic) inferential properties. The interested reader is referred to the appendix of van der Laan and Rose (2011) \cite{vanderLaan:2011}, or Schnitzer and Cefalu \cite{Schnitzer:2017}, for details on how the influence curve relates to inference in targeted maximum likelihood estimation.

\subsection{Inferential considerations for the three presented estimators}\label{sec:presentation}
Each of the above introduced estimators requires the estimation of conditional expectations which can be utilized with (parametric) regression models. Under the assumption that they are correctly specified this approach is valid. In a complex longitudinal setting such as ours, model mis-specification is a concern, and may yield a biased estimate of $\psi_T$.

An alternative to using regression models would be (supervised) machine learning, which is a loose term in the statistical literature referring to both traditional and modern estimation techniques, such as semi-parametric regression models, shrinkage estimators, boosting, random forests, support vector machines, and others.  For a given data set, and an a priori specified set of estimation techniques, the method minimizing the expected prediction error (as estimated by $k$-fold cross validation) can be chosen. As it is not clear beforehand which method this is, and whether a combination of learners might perform even better than a single one, one may use super learning. Super learning, which is also known as stacking \cite{Breiman:1996}, means considering a set of prediction algorithms (``learners''); Instead of choosing the algorithm with the smallest cross validation error, super learning chooses a weighted linear combination of different algorithms, that is the weighted combination which minimizes the $k$-fold cross validation error (for a given loss function, e.g. quadratic loss). This combination is linear, the weights are non-negative and sum up to one. The loss can be easily minimized with either quadratic programming or a non-negative least squares estimation approach, and is implemented in the $R$-package \texttt{SuperLearner} \cite{Polley:2017}. It can be shown that this weighted combination will perform asymptotically at least as well as the best algorithm, if not better \cite{vanderLaan:2008}. If none of the specified algorithms is a correctly specified parametric model, then the super learner will perform asymptotically as well as the oracle selector (which chooses the best weighted combination in the library), otherwise it will achieve an almost parametric rate of convergence (and will therefore perform reasonably well).
For all methods a major advantage of using super learning (rather than a single regression model) is that the probability of model mis-specification is reduced because distributional assumptions are relaxed and complex multivariate relationships can potentially be represented by the weighted combination of (potentially simple) learners. However, i) good prediction algorithms may still yield biased model estimates of the conditional expectations (in exchange for reduced variance) and ii) cross validation evaluates the expected prediction error with respect to the global fit of the conditional expectation, and not the scalar parameter $\psi_T$. TMLE incorporates the targeting step towards $\psi_T$ and can thus use super learning, whereas the (sequential) g-formula may not necessarily benefit from it.

The successful use of super learning requires a good and broad selection of prediction algorithms, and the ability to estimate these on the data. This may be challenging for longitudinal data with many time points and small sample size, as in our data. Section \ref{sec:MvM_machine_learning} gives a thorough discussion on the challenges and appropriate use of machine learning for data setups like ours.

\section{Data Analysis}\label{sec:data_analysis}
\subsection{Study setting, data and DAG}\label{sec:data_analysis_setting}

Our analysis includes data from 16 HIV treatment cohorts from Cote d'Ivoire, Burkina Faso, Ghana, Senegal, Togo, South Africa, Malawi, and Zimbabwe. All cohorts are part of the International epidemiology Databases to Evaluate AIDS (IeDEA) cohort collaboration, from both West Africa and Southern Africa \cite{Egger:2012, Ekouevi:2011}. HIV-positive children aged 12-59 months at enrolment, who acquired the virus before or at birth or during breast feeding, were included. To be eligible for the analysis, children had to have at least one visit before ART initiation, complete data at enrollment (baseline), and one follow-up visit. In total, $2604$ children were included in the analysis. We evaluated their data at $t^{\ast}=0,1,3,6,9,\ldots,30$ months after enrolment respectively. The follow-up time points refer to the intervals $(0,1.5)$, $[1.5,4.5)$, $[4.5,7.5)$, $[7.5,10.5)$, $[10.5,13.5)$, $[13.5,16.5)$, $[16.5,19.5)$, $[19.5,22.5)$, $[22.5,25.5)$, $[25.5,28.5)$, $[28.5,31.5)$ months. After 30 months of follow-up 1421 children were still alive and in care. Follow-up measurements, if available, refer to measurements closest to the middle of the interval. In our data $Y_{t^{\ast}}$ refers to height for age z-score (HAZ) at time $t^{\ast}$ (i.e. measured during the interval $(t^{\ast}-1,t^{\ast}]$). The binary treatment variable $A_{t^{\ast}}$ refers to whether ART was taken at time $t^{\ast}$ or not. Similarly, $C_{t^{\ast}}$ and $S_{t^{\ast}}$ describe whether a patient was censored (administratively or due to loss to follow-up), or died, at time $t^{\ast}$ respectively. Children were defined as being lost to follow-up if, at the time of database closure, they had not had contact with their health care facility for at least 9 months since their last recorded visit.  Time-dependent confounders affected by prior treatment refer to $\mathbf{L}_{t^{\ast}}=(L^1_{t^{\ast}}, L_{t^{\ast}}^2, L^3_{t^{\ast}}, L^4_{t^{\ast}})$ which are CD4 count, CD4\%, weight for age z-score (WAZ) which is a proxy for WHO stage \cite{Schomaker:2013}, i.e. a disease progression measure ranging from stage 1 (asymptomatic) to stage 4 (AIDS), and $L^4_{t^{\ast}} = Y_{t^{\ast}}$, $t^{\ast}<30$. Similarly, we write $L^{1m}_{t^{\ast}}, L_{t^{\ast}}^{2m}, L^{3m}_{t^{\ast}}, L^{4m}_{t^{\ast}}$ for indicator variables which specify if $L^s$ has been measured at time $t^{\ast}$ or not. They are needed to facilitate estimation step 3 for the g-formula because we don't want to intervene on measurement frequency but instead model irregular measurements as in reality. $\mathbf{L}_0$ refers to baseline values of CD4 count, CD4\%, WAZ, HAZ as well as sex, age, year of treatment initiation and region. Secondary outcome variables are $Y_{6}, Y_{12}, Y_{24}$.

Our assumptions about the structural relationships of these variables are summarized in the DAG in Figure \ref{figure:DAG}. It is known that (most recent) CD4 count and CD4\%, as well as WHO stage (approximated by WAZ \cite{Schomaker:2013}), i.e. $\mathbf{L}_{t^{\ast}}$ influence a clinician's decision about treatment assignment, affect the outcome and are affected by prior treatment. This is because previous HIV treatment guidelines have recommended using these disease severity markers to decide whether to prescribe ART or not \cite{Schomaker:2017}. Both the (sequential) g-formula and LTMLE can adjust for the type of confounding induced by $\mathbf{L}_{t^{\ast}}$. A child's growth depends on various factors, including ART \cite{Phutanakit:2012}, disease severity ($\mathbf{L}$) \cite{Schomaker:2016}, and unmeasured variables such as nutrition and socio-economic status (SES). It seems appropriate to assume that mortality depends predominantly on disease severity (and demographics) while the censoring process may depend on a multitude of factors. The time-ordering of the variables is based on the considerations of Section \ref{sec:MvM_structure}.

\subsection{Target parameter and interventions}
We consider the mean HAZ at $30$ months, under no censoring, for a given treatment rule $\bar{d}^j_{t^{\ast}}$, $j=1,...,4$, to be the main quantity of interest, that is $\psi_{30}=\mathbb{E}(Y_{30}^{\bar{d}^j_{t^{\ast}}})$. Secondary quantities are $\psi_6, \psi_{12}$, and $\psi_{24}$.

\clearpage
\begin{sidewaysfigure}[h!]
\begin{center}
\begin{tikzpicture}
\node[inner sep=0pt] (DAG) at (0,0)
    {\includegraphics[scale=0.775]{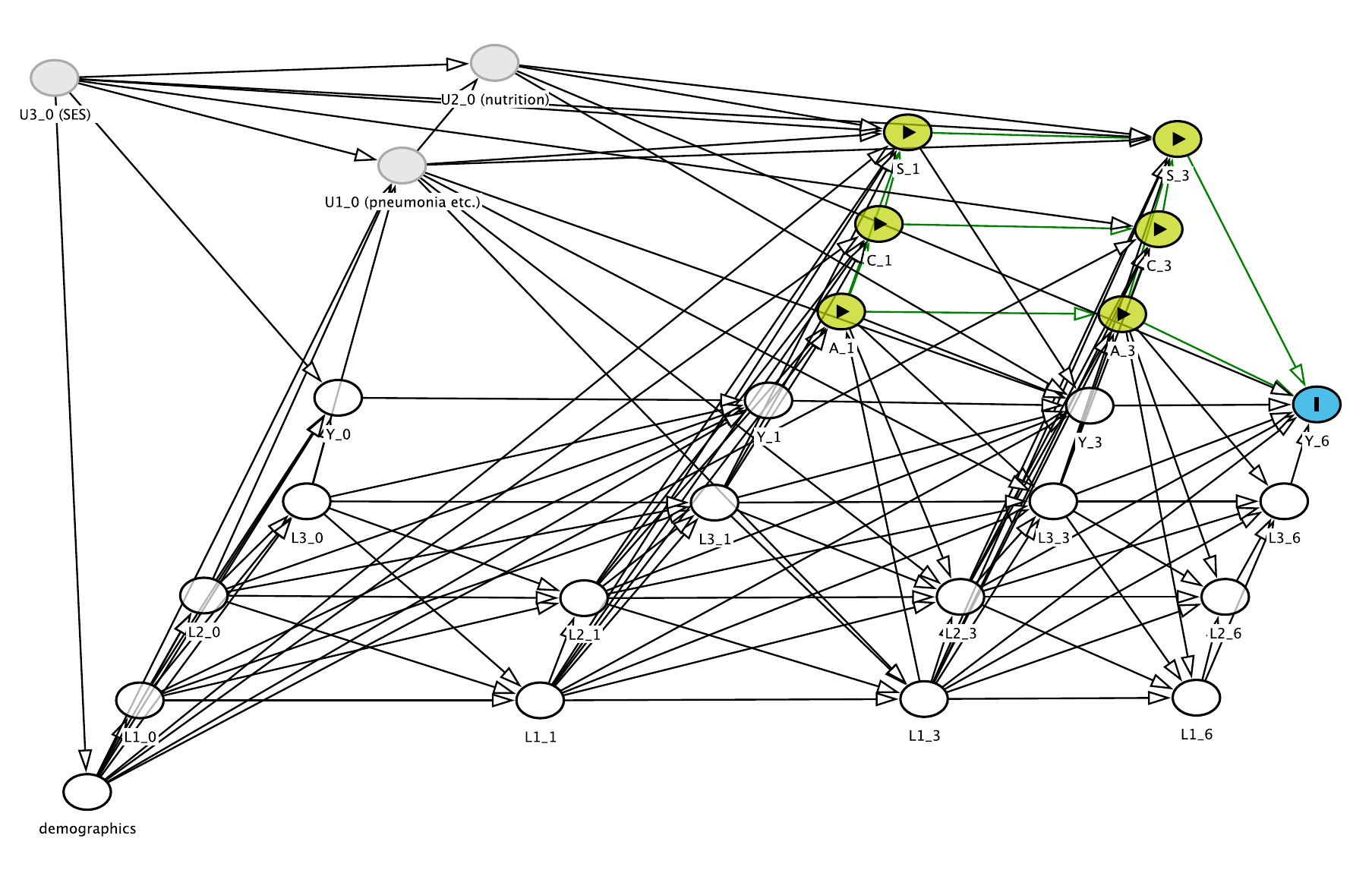}};
\node  (b1)  at (-12,-7.5)  {};
\node  (b2)  at (12,-7.5)  {};
\draw[->,very thick] (b1) -- (b2) node[midway,above] {time};
\end{tikzpicture}
\caption{DAG containing the structural assumptions about the data generating process. The target quantity is $\psi_6$, and relates to $Y_6$ which is represented by the $\mathbf{I}$ node. Intervention nodes (here: $A_1, A_3, C_1, C_3, S_1, S_3$) are represented by yellow nodes and the $\blacktriangleright$ symbol. Measured confounders are white, and unmeasured variables are grey. Causal paths for the intervention to the outcome are shown by green arrows.}
\label{figure:DAG}
\end{center}
\end{sidewaysfigure}
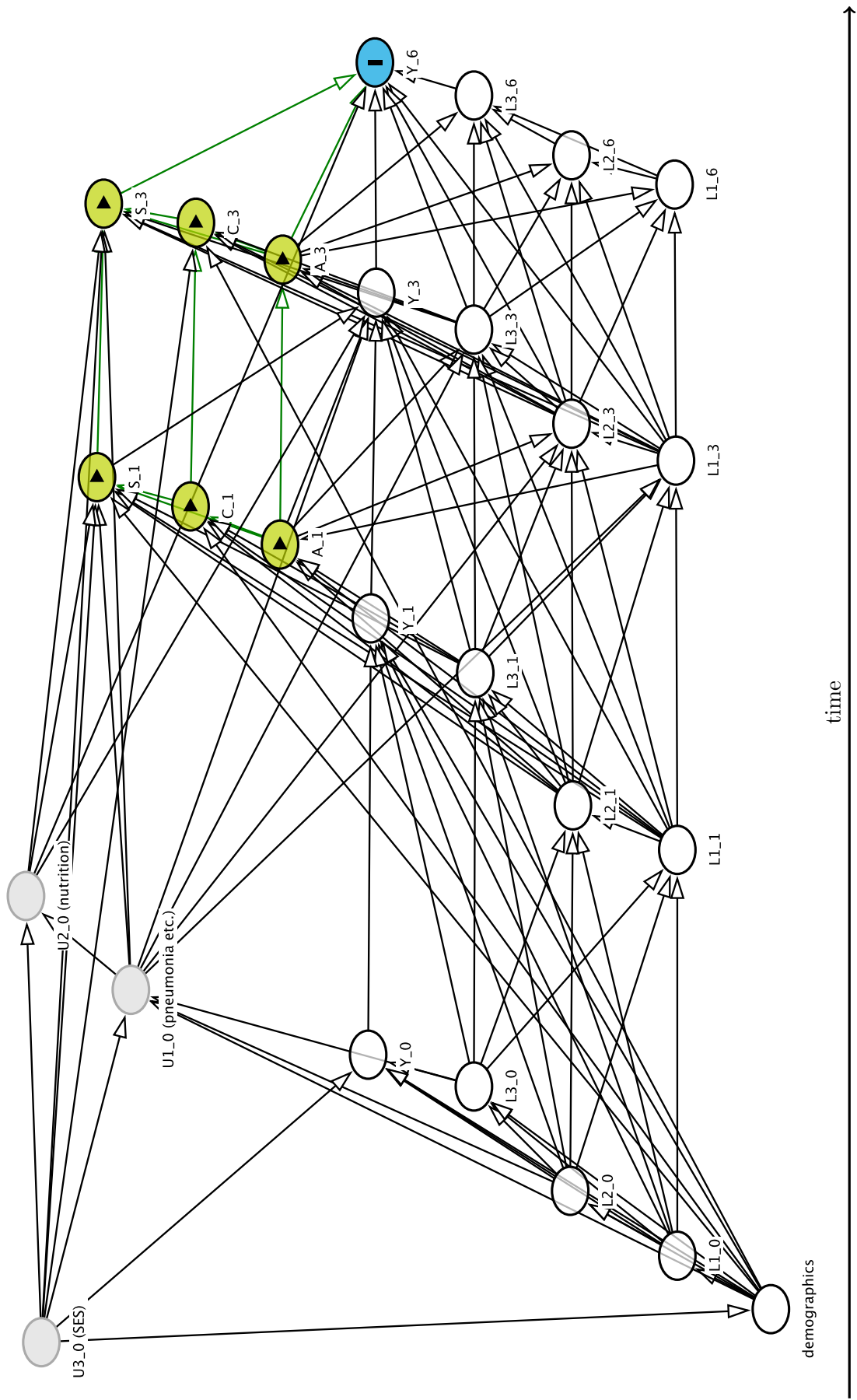
\clearpage

This target parameter is evaluated for the following four interventions, two of them static, two dynamic:

\begin{flalign*}
\bar{d}_{t^{\ast}}^{1} \phantom{(\text{CD4 count}_{t^{\ast}}, \text{CD4\%}_{t^{\ast}})} &= \,\, \left\{ c_{t^{\ast}} = 1; s_{t^{\ast}} = 1; a_{t^{\ast}}=1 \quad \text{for} \quad \forall t^{\ast}\right.&
\end{flalign*}
\begin{flalign*}
\bar{d}_{t^{\ast}}^{2} (\overline{\text{CD4 count}_{t^{\ast}}}, \overline{\text{CD4\%}_{t^{\ast}}}) &=\left\{ \begin{array}{cl}
               c_{t^{\ast}} = 1; s_{t^{\ast}} = 1; \quad a_{t^{\ast}}=1 & \quad \mbox{if} \quad \text{CD4 count}_{t^{\ast}}^{\bar{d}_{t^{\ast}}} < 750  \quad \text{or} \quad \text{CD4\%}^{\bar{d}_{t^{\ast}}}_{t^{\ast}} < 25\% \\ &\quad\quad\quad \text{or} \quad a_{t^{\ast}-1}=1 \\
               c_{t^{\ast}} = 1; s_{t^{\ast}} = 1; \quad a_{t^{\ast}}=0 &  \quad \mbox{otherwise}
               \end{array}
               \right. &
\end{flalign*}
\begin{flalign*}
\bar{d}_{t^{\ast}}^{3} (\overline{\text{CD4 count}_{t^{\ast}}}, \overline{\text{CD4\%}_{t^{\ast}}}) &= \left\{ \begin{array}{cl}
               c_{t^{\ast}} = 1; s_{t^{\ast}} = 1; \quad a_{t^{\ast}}=1 & \quad \mbox{if} \quad \text{CD4 count}_{t^{\ast}}^{\bar{d}_{t^{\ast}}} < 350  \quad \text{or} \quad \text{CD4\%}^{\bar{d}_{t^{\ast}}}_{t^{\ast}} < 15\% \\ &\quad\quad\quad \text{or} \quad a_{t^{\ast}-1}=1\\
               c_{t^{\ast}} = 1; s_{t^{\ast}} = 1; \quad a_{t^{\ast}}=0 &  \quad \mbox{otherwise}
               \end{array}
               \right. &
\end{flalign*}
\begin{flalign*}
\bar{d}_{t^{\ast}}^{4} \phantom{(\text{CD4 count}_{t^{\ast}}, \text{CD4\%}_{t^{\ast}})} &= \,\, \left\{ c_{t^{\ast}} = 1; s_{t^{\ast}} = 1; a_{t^{\ast}}=0 \quad \text{for} \quad \forall t^{\ast}\right.&
\end{flalign*}

The first intervention ($\bar{d}_{t^{\ast}}^{1}$) means immediate treatment initiation, the last one ($\bar{d}_{t^{\ast}}^{4}$) means no treatment initiation, and interventions two and three ($\bar{d}_{t^{\ast}}^{2}, \bar{d}_{t^{\ast}}^{3}$) mean deferring treatment until a particular CD4 threshold is reached. All interventions allow no censoring and death, and assume that a child stays on treatment once initiated.

\textit{Remark:} Note that the above target parameter is different from the quantities estimated in other analyses which are concerned with estimation of the effect of timing on treatment initiation on growth. These analyses evaluate the mean HAZ under a particular treatment strategy \textit{conditional} on either the observed subset of survivors at time $t^{\ast}-1$ (i.e. $S_{t^{\ast}-1}=1$, \cite{Puthanakit:2012}) or the subset of survivors that would have survived under the respective strategy (i.e. $S_{t^{\ast}-1}^{\bar{d}^{j}}=1$, \cite{Schomaker:2016, Schomaker:2017}). Conditioning on subsets of survivors is based on the rationale that one is typically interested in the benefits of treatment initiation beyond pure survival, i.e. general child development, including growth, among those children who are alive and can potentially be followed up. Unfortunately, conditioning on observed or counterfactual survival will lead to target parameters that can likely not be identified because conditional exchangeability may be violated. For example, by conditioning on the subset of observed survivors several paths, including $A_t \rightarrow S_t \leftarrow U \rightarrow Y_{t+1}$ would be opened as unmeasured variables $U$, such as past pneumonia or tuberculosis episodes, could potentially affect both survival as well as the outcome, i.e. one would face collider bias (see Figure \ref{figure:DAG}). Moreover, the population evaluated changes for each $t^{\ast}$, and also for each $\bar{d}^j_{t^{\ast}}$ if one conditions on $S_{t^{\ast}-1}^{\bar{d}^{j}}=1$; this makes comparisons difficult. Future research may evaluate alternative effect measures, such as the survivor average causal effect, which is partially identifiable and compares the outcome under two different treatment strategies, but only among the subpopulation that would have survived irrespective of which treatment strategy they were assigned to \cite{Eglestion:2007, Chiba:2011}.

\subsection{Estimation strategies}
 We estimate the target quantities $\psi_{30}$, $\psi_{24}$, $\psi_{12}$ and $\psi_6$ based on four approaches:
\begin{enumerate}[(i)]
\item using the g-formula as described in (\ref{eqn:gformula}) with a manual implementation which includes prior clinical prior knowledge (i.e. using only additive regression models),
\item using the sequential g-formula as described in (\ref{eqn:seq_g_formula}), using the \texttt{ltmle} package in $R$ \cite{Lendle:2017, Schwab:2016} with a data adaptive estimation approach,
\item using TMLE as described in Section \ref{sec:tmle} with a manual implementation which includes prior clinical knowledge (i.e. using only additive regression models),
\item using TMLE as described in Section \ref{sec:tmle}, using the \texttt{ltmle} package in $R$ \cite{Lendle:2017, Schwab:2016} with a data adaptive estimation approach.
\end{enumerate}
This choice essentially compares using standard software approaches (\texttt{ltmle}, existing and applicable super learning wrappers) with a manual implementation which uses only one learner (additive models), however one which is complex, not implemented in a wrapper, and incorporates prior clinical knowledge. Section \ref{sec:MvM_interactions} provides more details on the choice of these estimators.

\subsubsection{Model specification}

The implementation details for both LTMLE estimators are given in appendices \ref{sec:appendix_ltmle_aut} and \ref{sec:appendix_ltmle_man}, the supplementary material, and are discussed further in Section \ref{sec:MvM}. Briefly, for the automated implementations, i.e. estimators (ii) and (iv), we use the following specifications: first, we do not specify the regression models for outcome, treatment, and censoring parametrically, but use super learning instead.  The candidate learners were the arithmetic mean, (generalized linear) regression models with all main terms [GLM], GLMs based on an EM-algorithm-Bayesian model fitting, GLMs chosen by stepwise variable selection with Akaikes Information Criterion [AIC], GLMs containing interaction terms, as well as additive models -- i.e. those learners which could be successfully fitted on our data.

Our manual implementation of the g-formula followed Schomaker et al. \cite[Appendix]{Schomaker:2016}. Briefly, we used semi-parametric additive linear and logistic regression models to model the conditional densities in (\ref{eqn:gformula}) and (\ref{eqn:gformula2}); To allow for flexible disease progression depending on how sick children are when they present at their first visit, interactions of baseline characteristics (represented in categories) with all other variables were added. Depending on the functional form of the covariates these interactions were either linear or non-linear (using penalized splines). Model selection by means of generalized cross validation was used to improve the bias-variance tradeoff in these models. All models implicitly assume that time-dependent risk factors measured before or on time $t^{\ast}-2$ do not predict the respective outcome. Note that the same approach of model building was used for our manual LTMLE estimator (iii). Sections  \ref{sec:MvM_machine_learning} and \ref{sec:MvM_interactions} give a thorough discussion of the issue of model choice.

\subsubsection{LTMLE implementation}
Both the automated and the manual implementations of the LTMLE algorithm do not use linear regression models to model the outcome, HAZ; rather, the data are transformed to lie between $[0,1]$ and a Quasi-binomial model is used \cite{McCullagh:1989}. This is the way the \texttt{ltmle} package deals with continuous outcomes with the aim of improving stability, particularly in the context of (near) positivity violations \cite{Gruber:2010}.

The clever covariate as defined in step 3 in Section \ref{sec:tmle} is included in the regression model not as a covariate, but as a weight \cite{Kreif:2017}. This is another valid targeted maximum likelihood estimator, based on a refined loss function (and parametric submodel), and can improve stability \cite{Robins:2007, Tran:2019}. This approach is again part of the \texttt{ltmle} implementation.

Moreover, the denominator of the clever covariate (\ref{eqn:clever_covariate}) is bounded in the sense that values $<g$ are truncated at $g$. The automated implementation uses the default of $g=0.01$, the manual implementation a stricter version of $g=0.05$; see Section \ref{sec:MvM_gbounds} for a thorough discussion on truncation of the clever covariate.

\subsubsection{Other details}

Our automated implementation of the sequential g-formula, i.e. estimator (ii), follows the automated implementation of LTMLE, except that step 3 in Section \ref{sec:tmle} is not part of the inference procedure.

All four methods use the square root of CD4 count to work with a more symmetric variable.

\subsection{Results}
\begin{sidewaysfigure}[h!]
\begin{center}
\includegraphics[scale=0.65]{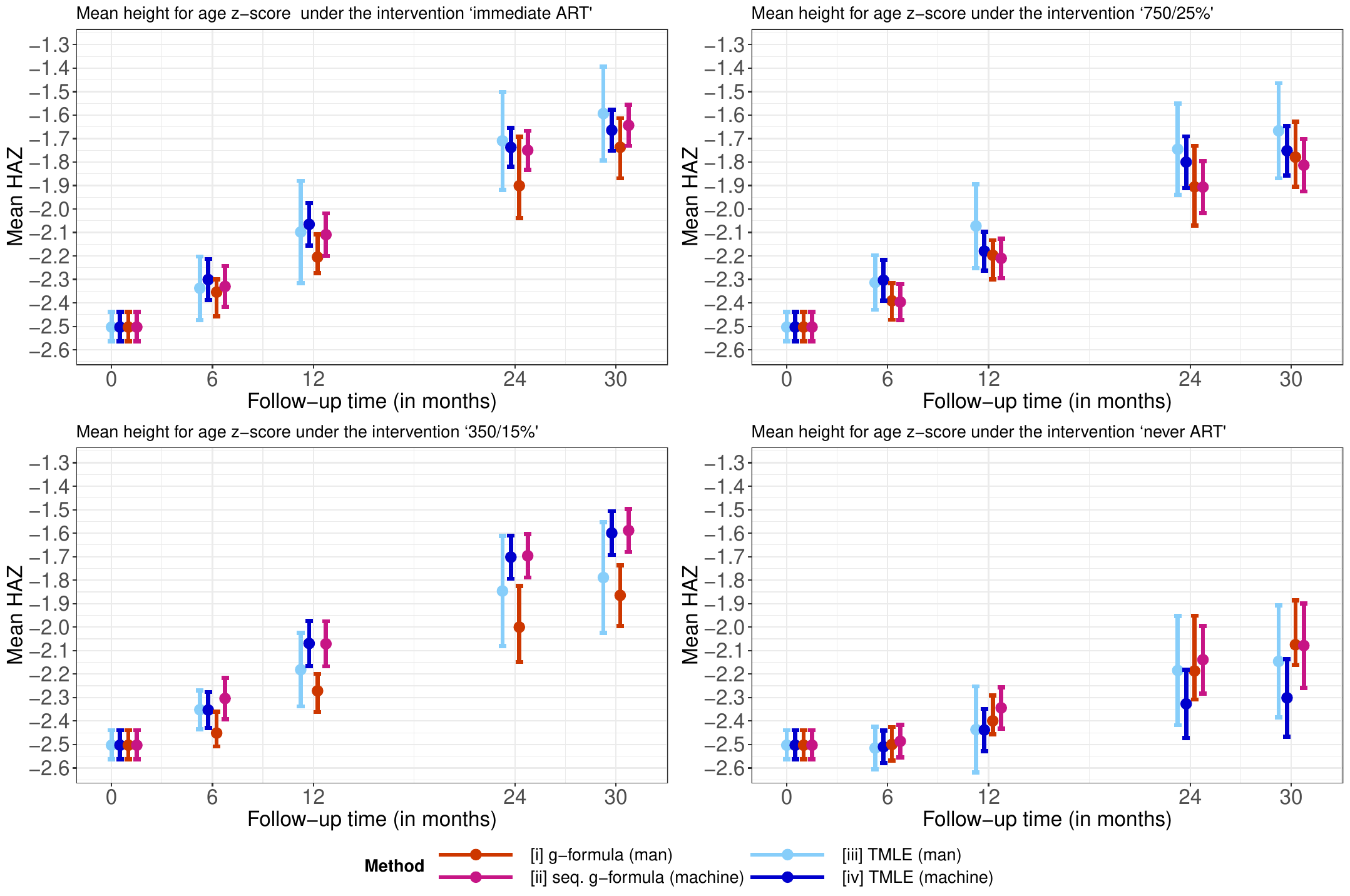}
\caption{Mean HAZ for different interventions, time points, and methods.}
\label{figure:interval_combined}
\end{center}
\end{sidewaysfigure}

Figure \ref{figure:interval_combined} summarizes the results of the analyses, i.e the point and interval estimates for $\psi_6$, $\psi_{12}$, $\psi_{24}$, $\psi_{30}$.

For immediate ART initiation, and the dynamic intervention of starting ART if either CD4 count $<750$ or CD4\% $<25$\%, the point estimates of all estimators were generally similar (Figure \ref{figure:interval_combined}, top panel). Deviations after 30 months of follow-up were not much greater than $0.1$.

For the second dynamic intervention (Figure \ref{figure:interval_combined}, bottom left) estimates from the manual implementations (i.e. estimators (i) and (iii)) where generally substantially lower than the estimates from the automated implementations. One of the most interesting results can be seen for the intervention ``never ART'': here, automated estimation via LTMLE deviated by approximately -0.2 from all other methods (Figure \ref{figure:interval_combined}, bottom right).

For all results confidence intervals were widest for our manual implementation of LTMLE, followed by our manual implementation of the g-formula. There is a tendency for all estimators to suggest better outcomes with earlier treatment initiation; however, the automated implementations are less clear about this for the intervention that uses the 350/15\% threshold. Also, the automated TMLE procedure suggests more severe consequences of withholding antiretroviral treatment (``never ART'') than the other estimators.

In summary, the four methods produce slightly different results. The manual versions of the g-formula and LTMLE suggest that earlier treatment leads to better height gain. Our automated LTMLE procedure deviates from these results by considering the 350/15\% intervention more positive and the never ART rule more negative.

Interestingly, the automated implementations of the sequential g-formula and LTMLE hardly differ for the first three interventions of interest, which is reassuring, but differ substantially for the fourth intervention of no antiretroviral treatment initiation. This could relate to positivity violations, as outlined in Section \ref{sec:MvM_gbounds} and the simulation studies.

Our manual implementations of the g-formula and TMLE suggest broadly similar growth differences between the different intervention strategies as Schomaker et al. \cite{Schomaker:2016}, who implemented a similar analysis on a much bigger dataset using the g-formula with good performance under the natural course scenario (see also Section \ref{sec:MvM_natural_course}), but with an emphasis on mortality estimates.

For all interventions and time points, super learning made extensive use of additive models in the Q-models (weights between 0.44 and 0.85), and moderate use in the g-models (weights between 0.18 and 0.65). It is worth pointing out that for the g-models, the arithmetic mean, which is a simple learner, was important with average weights between 0.15 and 0.61.

\section{Practical Challenges in the Data Analysis and Diagnostics}\label{sec:MvM}
Below, in Sections \ref{sec:MvM_sample_size}-\ref{sec:MvM_interactions}, we highlight the general practical challenges of our data analysis which are relevant to other longitudinal settings as well. Sections \ref{sec:MvM_natural_course} and \ref{sec:MvM_gbounds} discuss possible diagnostic approaches for both the g-formula and LTMLE.

\subsection{Challenge: long follow-up and declining sample size}\label{sec:MvM_sample_size}
All 4 estimators under consideration require regression models (or data adaptive approaches) to be fitted on all subjects who are alive and uncensored at time $t$. It is evident that the longer the follow-up, the fewer subjects that can be used to fit the respective models, e.g. after 30 months of follow-up only 1421 patients remain.

\clearpage
\clearpage

As we discuss below, this can be a problem as all estimators rely on correct model specifications and with smaller sample sizes it may not be possible to use more complicated methods, such as additive models with non-linear interactions or advanced machine learning algorithms. Moreover, this problem can potentially become even more severe in step 3 of the LTMLE algorithm presented in Section \ref{sec:tmle_implementation}: here, the models are estimated on the subset of patients alive, uncensored \textit{and following treatment} $\bar{A}_t=\bar{d}_t$.

In our data, after 30 months of follow-up, we have $371$ (14.2\%) patients alive and uncensored following rule $\bar{d}^1_{t^{\ast}}$, $396$ (15.2\%) patients alive and uncensored following rule $\bar{d}^2_{t^{\ast}}$, $505$ (19.4\%) patients alive and uncensored following rule $\bar{d}^3_{t^{\ast}}$, and $292$ (11.2\%) patients alive and uncensored following rule $\bar{d}^4_{t^{\ast}}$. This is the available sample size for the LTMLE estimator during its updating step. Note that after 30 months of follow-up we have evaluated $3$ confounders, the outcome, as well as treatment at 11 different time points. Together with the baseline confounders this makes about 60 covariates which are potentially relevant to the respective regression models at time $30$. With adding non-linear relationships, for example via penalized splines, the number of parameters that needs to be estimated can therefore be close to $n$. The consequence is that complex estimation procedures, which may be needed for data adaptive approaches, may either fail or yield unstable solutions.

\subsection{Challenge: making structural temporal assumptions}\label{sec:MvM_structure}
In general, it is required to make assumptions about the time-ordering of the data. For our data analysis the time-ordering assumptions can be visualized as follows:

\begin{center}
\begin{tikzpicture}
\node  (b1)  at (0,0) [circle,draw=black,fill=black!10!white,minimum size=1cm] {$\mathbf{L}_0$};
\node  (b2)  at (1.5,0) [circle,draw=black,fill=black!10!white,minimum size=1cm] {$L_1^1$};
\node  (b3)  at (3,0) [circle,draw=black,fill=black!10!white,minimum size=1cm] {$L_1^2$};
\node  (b4)  at (4.5,0) [circle,draw=black,fill=black!10!white,minimum size=1cm] {$L_1^3$};
\node  (b5)  at (6,0) [circle,draw=black,fill=black!0!white,minimum size=1cm] {$Y_1$};
\node  (b6)  at (7.5,0) [circle,draw=black,fill=black!30!white,minimum size=1cm] {$A_1$};
\node  (b7)  at (9,0) [circle,draw=black,fill=black!30!white,minimum size=1cm] {$C_1$};
\node  (b8)  at (10.5,0) [circle,draw=black,fill=black!20!white,minimum size=1cm] {$S_1$};
\draw[->, thick] (b1) to (b2);
\draw[->, thick] (b2) to (b3);
\draw[->, thick] (b3) to (b4);
\draw[->, thick] (b4) to (b5);
\draw[->, thick] (b5) to (b6);
\draw[->, thick] (b6) to (b7);
\draw[->, thick] (b7) to (b8);
\node  (c1)  at (0,-1.5) [fill=black!0!white,minimum size=1cm] {$...$};
\node  (c2)  at (1.5,-1.5) [circle,draw=black,fill=black!10!white,minimum size=1cm] {$L_{30}^1$};
\node  (c3)  at (3,-1.5) [circle,draw=black,fill=black!10!white,minimum size=1cm] {$L_{30}^2$};
\node  (c4)  at (4.5,-1.5) [circle,draw=black,fill=black!10!white,minimum size=1cm] {$L_{30}^3$};
\node  (c5)  at (6,-1.5) [circle,draw=black,fill=black!0!white,minimum size=1cm] {$Y_{30}$};
\draw[->, thick] (c1) to (c2);
\draw[->, thick] (c2) to (c3);
\draw[->, thick] (c3) to (c4);
\draw[->, thick] (c4) to (c5);
\end{tikzpicture}
\end{center}

Note that $\mathbf{L}_0$ includes $Y_0$, and that we do not explicitly specify $A_0$, $C_0$, and $S_0$ here, as by definition everyone is uncensored, untreated and alive at $t^{\ast}=0$. Equation (\ref{eqn:gformula2}) implies that the ordering of the time-dependent confounders $\mathbf{L}_{t^{\ast}}$ can possibly affect the results of the g-formula: the confounder distribution which comes last, i.e the one referring to $L_{t^{\ast}}^3$, can be informed by the other confounders measured at the same time point, $L_{t^{\ast}}^{1,2}$, but not vice versa. While all factorizations of $\mathbf{L}_{t^{\ast}}$ are valid, they may yield somewhat different answers in a finite sample. It is clear that g-formula analyses benefit from sensitivity analyses with respect to the ordering of the time-dependent confounders. However, this may be very time-consuming or even unfeasible given the additional burden of bootstrapping. For LTMLE estimators the ordering of the $L^s$ does \textit{not} play a role (because no models for $L_{t^{\ast}}^s$ are required), and neither is bootstrapping needed -- which shows the method's benefits in this regard. Nevertheless, the assumed time ordering still plays a role for LTMLE, as the modeled iterated outcome and treatment and censoring mechanisms may only include those variables which are assumed to be measured prior to the respective variable.

Given the above considerations, the obvious choice for $L^3$ in our data is WAZ, as it is a strong proxy for disease severity and physical development \cite{Schomaker:2013}. At each time point WAZ can be informed by CD4 count and CD4\% at the same time point, but not vice versa. Since, for children, CD4\% is a more accurate measure of disease severity than CD4 count it makes sense to treat the former as $L^2$ and the latter as $L^1$. Treatment decisions ($A$) can potentially be informed by laboratory measurements ($L^1, L^2$) and clinic measurements ($L^3, Y$) which we have acknowledged in our time ordering. Since death and censoring can happen until the end of an interval, and a patient may still experience measurements before these events, we assume $S$ and $C$ to be measured at the end of each interval.

\subsection{Challenge: machine learning in the light of long follow-up and declining sample size}\label{sec:MvM_machine_learning}
The successful use of super learning for LTMLE estimators has been shown repeatedly in the literature \cite{Neugebauer:2014,vanderLaan:2011}. However, the problem for our motivating example is that particularly for longer follow-up times most of the available machine learning algorithms failed. In fact, we were only successful in using the learners described in Section \ref{sec:data_analysis}. All other algorithms provided by the \texttt{SuperLearner} package failed. One could argue that under such circumstances the use of super learning is limited, unless one is able to and willing to write custom-made wrappers.

Reasons for failure were manifold: (i) memory problems because of the use of complex fitting algorithms in the context of large covariate data (ii) non-convergence of fitting algorithms because of $p \approx n$ [as motivated in Section \ref{sec:MvM_sample_size}] (iii) sparse covariate data (iv) un-identified problems of the fitting algorithm (v) inability to use existing wrappers for Quasi-binomial models \cite{McCullagh:1989}, which are used in \texttt{ltmle} for bounded and transformed continuous outcomes and (vi) others. Note that we need predictions from 3 prediction models/algorithms (one in step 1 and two in step 3c) for 11 time points, and for 4 interventions. Each time we work with a different subset of data with different covariate combinations. If problems occur for one model, potentially because of the specific sub-sample of data, super learning can fail as a whole. To tackle this problem one could pool models over time points \cite{Schnitzer:2016}, though this approach is not implemented in \texttt{ltmle}.

\subsection{Challenge: incorporating prior clinical knowledge using non-linear interactions}\label{sec:MvM_interactions}
A clinical hypothesis in our data is that patients who are sicker at presentation will have a different disease trajectory from patients who are healthier at presentation. Children who arrive with severe disease will not be able to recover as fast as healthy children and will also not be able to reach the same CD4 count levels \cite{Schomaker:2016}. A model describing this relationship could be of the form
\begin{eqnarray*}
g(E(Y_t)) &=& \mathbf{X\beta} + f_1(L_{t}^1,L_{1}^3) + f_2(L_{t}^2,L_{1}^3) + f_3(L_{t}^3,L_{1}^3),
\end{eqnarray*}
or similar, where $g(\cdot)$ represents a link function (i.e. the identity link if $Y$ refers to HAZ), $\mathbf{X\beta}$ refers to the parametric part of the model (e.g. binary baseline variables), and $f_j$ are unspecified general smoothing functions.  The above model is a generalized additive model in its wider sense \cite{Wood:2006}. No matter whether using the g-formula or an LTMLE estimator it is important to deal with many multiple non-linear interactions at the same time. The $R$ library \texttt{mgcv} can fit models as specified above, and this is the reason we use them in our manual implementations, i.e. estimators (i) and (iii). Using additive models in this context requires stepwise manual inclusion of non-linear interactions, if they improve the prediction error in terms of generalized cross validation, because inclusion of all possible interactions will make the fitting process too complex to converge and also results in models with unacceptably high variance.

Note that the automated estimation procedures (ii) and (iv) include the \texttt{gam} wrapper provided by the \texttt{SuperLearner} package. However, this wrapper doesn't use the \texttt{mgcv} implementation described above but another one from package \texttt{gam}. While the latter has advantages in that it uses the simple backfitting algorithm \cite{Breimann:1985} for estimation, and could therefore be regarded as non-parametric rather than semi-parametric, it lacks a flexible approach of dealing with non-linear interactions, i.e. fitting smooth functions from a first variable that vary with respect to the categories of a second variable. Moreover, it does not use automated smoothing parameter selection using cross validation but rather needs pre-specified knowledge of the degrees of freedom of the model. Also, the default wrapper implicitly assumes no (non-linear) interactions. This fact can be irrelevant for super learning if enough learners are available, such that the combined weighted super learner contains the predictive information related to the interactions; however, if the number of (pre-implemented) learners that can be fitted is limited, as in our data analysis, the question arises whether it may be preferable to proceed with a manual implementation rather than a restricted automated one. This question will be answered later in this paper. Of course, it is also possible for the user to write their own wrapper function, based on \texttt{mgcv}, with model selection with respect to non-linear interactions, though this would be rather computationally intensive.

\subsection{Diagnostics for the parametric g-formula: the natural course scenario}\label{sec:MvM_natural_course}
Applying the g-formula assumes that the densities in (\ref{eqn:gformula}) are estimated consistently. A necessary, but not sufficient, condition for correct specification is that the g-formula re-produces the raw data (at least approximately) if the treatment rule $\bar{d}_{t^{\ast}}$ relates to the real treatment assignment process in the data. More specifically, the ``natural course scenario'' evaluates the stochastic treatment rule where treatment assignment probabilities are estimated from $\mathbb{P}({A}_{t^{\ast}}=a_{t^{\ast}}|\bar{A}_{t^{\ast}-1}, \mathbf{\bar{L}}_{t^{\ast}})$ and censoring and death is not intervened upon but modeled via $\mathbb{P}({C}_{t^{\ast}}=1|\bar{A}_{t^{\ast}}, \mathbf{\bar{L}}_{t^{\ast}})$ and $\mathbb{P}({S}_{t^{\ast}}=1|\bar{A}_{t^{\ast}}, \mathbf{\bar{L}}_{t^{\ast}})$. All models are fitted on those alive and uncensored. Then, $\hat{\psi}_T = \frac{1}{n} \sum_{i=1}^n \hat{Y}^{\bar{d}^{j}_{t^{\ast}}}_{T,i}$ (and also $\hat{\bar{L}}^{s,\bar{d}^{j}_{t^{\ast}}}_{t^{\ast}}$) under this stochastic treatment rule, and estimated by the g-formula as described in Section \ref{sec:gformula_implementation}, can be compared to the mean HAZ at the end of follow-up in the data, and should approximately match. Reporting the natural course scenario is a standard procedure in g-formula analyses \cite{Westreich:2012, Schomaker:2016, Young:2011}.  We note however that the approach used in the natural course scenario evaluates a target quantity conditional on survival, as opposed to the target quantity defined in Section \ref{sec:framework} where survival is intervened upon; thus caution should be exercised when interpreting the below results.

\begin{figure}[ht!]
\begin{center}
\subfloat[observed mean HAZ compared to the counterfactual mean HAZ under the natural course scenario for different time points]{\label{figure:natural_course_a}\includegraphics[scale=0.375]{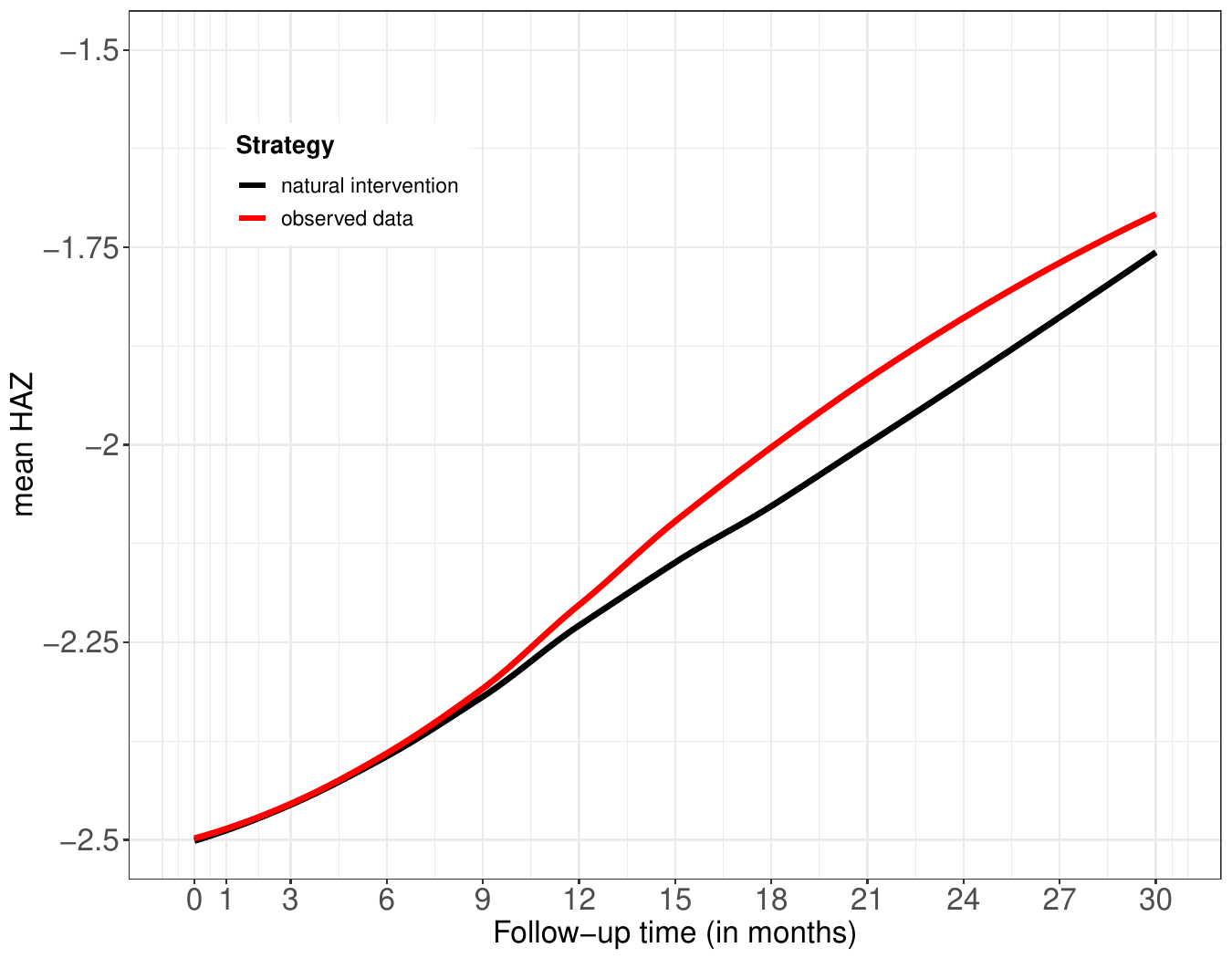}}\hspace*{0.2cm}
\subfloat[difference of the mean HAZ under the natural course scenario and the mean HAZ from the observed data (grey shaded area)]{\label{figure:natural_course_b}\includegraphics[scale=0.375]{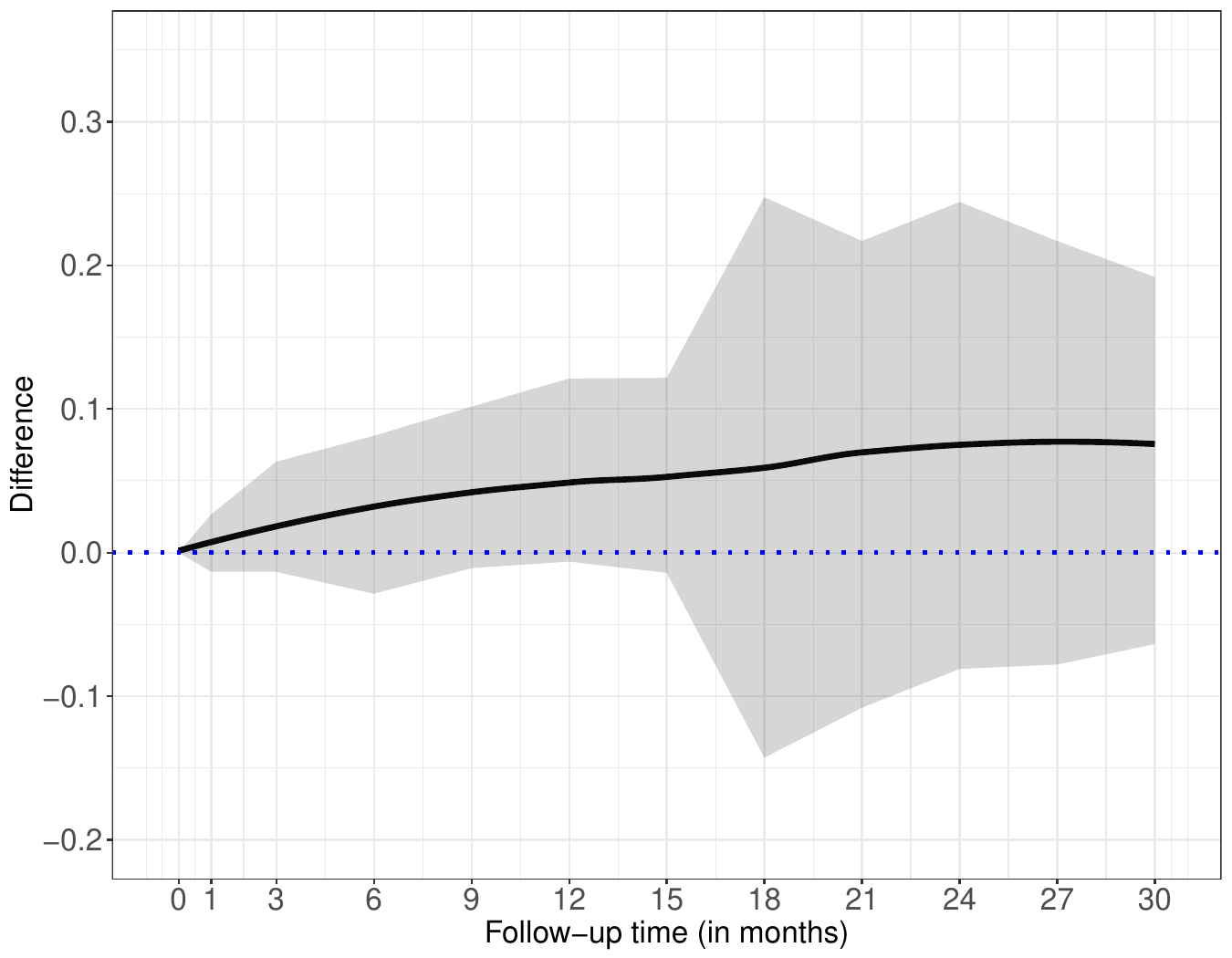}}
\caption{Diagnostic for the g-formula: natural course scenario} \label{figure:natural_course}
\end{center}
\end{figure}

Figure \ref{figure:natural_course} shows the natural course scenario for our data. The mean HAZ at the end of follow-up for both the raw data (``observed data'') and the data produced by the g-formula under the natural course scenario (``natural intervention'') are similar up to 1 year of follow-up, after which some deviation can be observed. The 95\% confidence interval of the difference (Figure \ref{figure:natural_course_b}) suggests that the difference may not be different from $0$, but some caution with respect to model mis-specification for our manual implementations remains.

\subsection{Diagnostics for LTMLE: inverse probabilities and truncation to detect positivity problems}\label{sec:MvM_gbounds}
As opposed to the g-formula, LTMLE estimators require models for the treatment and censoring mechanism. This mechanism is reflected in the clever covariate used in step 3\textit{c} of the LTMLE algorithm. While the LTMLE estimator is less sensitive than traditional inverse probability of treatment estimators \cite{Petersen:2012} it nevertheless includes inverse probability weights in its fitting process which may lead to more variable estimates when the weights become more variable \cite{Schnitzer:2017}.

For our motivating data example we have 20 treatment models relating to both the treatment and censoring mechanism of the time points. Ideally the fitted \textit{cumulative} probabilities are not too small because small estimated probabilities may indicate near-positivity violations; i.e. if there are covariate strata where the (estimated) probability of (not) receiving treatment is close to zero, for example when an extremely sick patient has not been treated although this would have been expected according to the data of similar patients. Practical positivity violations occur easily when dealing with many continuous confounders as in our data; but if the data is still ``rich'' enough both the g-formula and the LTMLE estimator may be able to deal with it and extrapolate based on the available data and estimated models \cite{Petersen:2012}.

One way to deal with small probabilities is to truncate them at a lower bound $g$. This reduces variability and improves stability, however at the cost of bias if the truncation is too severe \cite{Gruber:2010}.

Table \ref{tab:cc} summarizes mean clever covariates (i.e. cumulative inverse probabilities until 30 months of follow-up) for different model specifications and interventions. Their mean should be about 1. More importantly, the percentage of truncated probabilities (here, [cumulative] probabilities $<$ 0.01 are truncated), should be as small as possible.

\begin{table}[ht!]
\caption{Mean clever covariate for LTMLE estimators (at $t^{\ast}=30$) corresponding to different interventions and model specifications. The percentage of truncated values refers to the proportion of probabilities $<$ 1\%.}
\label{tab:cc}
\begin{center}
\begin{tabular}{r|c|ccc|ccc}
&& \multicolumn{3}{c|}{\%truncated}& \multicolumn{3}{c}{mean clever covariate}\\
Intervention & N & SL & GAM & GLM & SL & GAM & GLM \\
\hline
no ART          & 292&	0.7&68.5 &100	 &0.97   &8.96  &11.21    \\
350/15\%        & 505&	0  &32.9 &45.9	 &1.21   &9.02	&9.32    \\
750/25\%        & 396&	0.3&18.2 &11.6	 &0.88   &4.96  &2.28    \\
immediate ART   & 371&	0  &15.6 &0  	 &0.83   &4.30	&0.53    \\
\end{tabular}
\end{center}
\end{table}

GLM's which do not use any transformations or interactions and include covariate history up to time $t-2$ yield highly variable results, i.e. the mean clever covariate ranges from 0.53 to 11.21 and between 0\% and 100\% of truncated observations. Our manual implementation of TMLE [estimator (iii)], which uses additive models with non-linear interactions (``GAM''), leads to unstable weights with means that are far too high ($>4$) and many small probabilities which have been truncated. Using super learning (``SL'') yields good mean clever covariates, but still 0.7\% truncated values for the intervention of `no ART'.

Table \ref{tab:estimates_probabilities} lists point and confidence interval estimates of LTMLE estimators ($t=30$), as well as the corresponding mean clever covariate and truncation percentages, for different model specifications for the potentially problematic intervention $\bar{d}^4$.

\begin{table}[ht!]
\caption{Point and confidence interval estimates of LTMLE estimators ($t^{\ast}=30$), as well as the corresponding mean clever covariate, for different model specifications and truncation definitions. Results are provided for the intervention ``no ART''.}
\label{tab:estimates_probabilities}
\begin{center}
\begin{tabular}{r|cc|cc|cc}
\multicolumn{1}{c|}{}& \multicolumn{2}{c|}{Estimates} & \multicolumn{2}{|c}{mean CC} & \multicolumn{2}{|c}{\% truncated}\\
LTMLE  & g=0.01 & g=0.05 & g=0.01 & g=0.05& g=0.01 & g=0.05\\
\hline
GLM          &-2.05 (-3.53; -0.57)& -2.05 (-2.36; -1.75)  &11.21 & 2.24  &100   &100\\
GAM          &-2.19 (-3.19; -1.18)& -2.15 (-2.38; -1.91)  &8.96  & 2.14  &68.5  &90.8\\
SL           &-2.30 (-2.47; -2.14)& -2.25 (-2.35; -2.14)  &0.97  & 0.76  &0.7   &6.2\\ \end{tabular}
\end{center}
\end{table}

It can be seen that model specification has a considerable impact on the results. While truncation doesn't affect the point estimates much, it is worth noting that an improved mean clever covariate comes at the cost of truncation levels way above 5\%, even with super learning. This raises the question whether positivity violations are so severe that one wouldn't trust any of the provided estimates for the fourth intervention of interest.
Interestingly, our manual implementation of LTMLE (using additive models) yields estimates similar to the g-formula --  in spite of the poor weights. This can be explained as follows: the cumulative probabilities are so small that most of them get truncated and the clever covariate can therefore hardly update the initial g-formula estimate in step 3 of the TMLE algorithm.

\section{Simulation Study}\label{sec:simulation}
The following simulation is meant to investigate under which conditions effect identification with LTMLE may be difficult, i.e. to what degree small samples, limited sets of learners, overuse of truncation and intervention definitions may affect its performance. Based on our data analysis and the discussions in Section \ref{sec:MvM} we generate longitudinal data ($t=0,1,\ldots,12$) for 3 time-dependent confounders, as well as baseline data for 7 variables, using structural equations and the $R$-package \texttt{simcausal} \cite{Sofrygin:2016}. Baseline data includes region, sex, age, CD4 count, CD4\%, WAZ and HAZ respectively ($\mathbf{L}_0$). Follow-up data refers to CD4 count, CD4\%, WAZ and HAZ ($L^1_t$, $L^2_t$, $L^3_t$, $Y_t$), as well as antiretroviral treatment ($A_t$) and censoring ($C_t$). For simplicity, no deaths are assumed. The data generating mechanism contains interactions and non-linear relationships and is described in detail in Appendix \ref{sec:appendix_data_generating}. It leads to the following baseline values: region A = 75.5\%; male sex = 51.2\%; mean age  = 3.0 years; mean CD4 count = 672.5 cells/microlitre; mean CD4\% = 15.5\%; mean WAZ = -1.5; mean HAZ = -2.5. At $t=12$ the mean of CD4 count, CD4\%, WAZ and HAZ are 1092 cells/microlitre, 27.2\%, -0.8, -1.5 respectively. The target quantities $\psi_{j,t}$ are defined as the expected value of $Y$ at time $t$, $t \in \{6,12\}$, under no censoring, for a given treatment rule $\bar{d}^j$, where
\begin{flalign*}
\bar{d}_{t}^{1}  &= \,\, \left\{ c_{t} = 1; \quad \text{and} \quad a_{t}=1 \quad \text{for} \quad \forall t\right.&
\end{flalign*}
\begin{flalign*}
\bar{d}_{t}^{2}  &=\left\{ \begin{array}{cl}
               c_{t} = 1; \quad \text{and} \quad a_{t}=1 & \quad \mbox{if} \quad \text{CD4 count}_{t}^{\bar{d}_t^2} < 750  \quad \text{or} \quad \text{CD4\%}^{\bar{d}_t^2}_{t} < 25\%  \quad \text{or} \quad \text{WAZ}^{\bar{d}_t^2}_{t} < -2 \\& \quad \text{or} \quad a_{t-1}=1\\
               c_{t} = 1; \quad \text{and} \quad a_{t}=0 &  \quad \mbox{otherwise}
               \end{array}
               \right. &
\end{flalign*}
\begin{flalign*}
\bar{d}_{t}^{3} &= \left\{ \begin{array}{cl}
               c_{t} = 1; \quad \text{and} \quad a_{t}=1 & \quad \mbox{if} \quad \text{CD4 count}_{t}^{\bar{d}_t^3} < 350  \quad \text{or} \quad \text{CD4\%}^{\bar{d}_t^3}_{t} < 15\% \quad \text{or} \quad \text{WAZ}^{\bar{d}_t^3}_{t} < -2 \quad \\& \text{or} \quad a_{t-1}=1\\
               c_{t} = 1; \quad \text{and} \quad a_{t}=0 &  \quad \mbox{otherwise}
               \end{array}
               \right. &
\end{flalign*}
\begin{flalign*}
\bar{d}_{t}^{4} &= \,\, \left\{ c_{t} = 1; \quad \text{and} \quad a_{t}=0 \quad \text{for} \quad \forall t\right.&
\end{flalign*}

The true target parameters are $\psi_{1,6}=-1.804$, $\psi_{2,6}=-1.850$, $\psi_{3,6}=-2.028$, $\psi_{4,6}=-2.452$, $\psi_{1,12}=-1.029$, $\psi_{2,12}=-1.119$, $\psi_{3,12}=-1.466$, and $\psi_{4,12}=-2.447$.

For $\mathcal{R}=1000$ simulation runs we estimate the above parameters using \texttt{ltmle} under the following specifications: $n \in \{200,600,1000\}$; truncation levels $g \in \{0.01, 0.025, 0.04\}$ and three different sets of learners. Learner 1 consists of simple GLM's, learner 2 adds both the arithmetic mean and GLM's with interactions to the set of learners, and learner 3 contains additionally GLM's selected by AIC, standard additive models (no interactions, package \texttt{gam}), as well as GLM's based on an EM-algorithm-Bayesian model fitting. For computational feasibility, all fitted models included only variables measured at time $t$, $t-1$, and $0$; in line with the data generating process.

The average useable size of the sample at the end of follow-up (i.e. excluding censored patients and only including those who followed the respective treatment rule) varied between 39 (intervention $\bar{d}_t^4$) and 79 (intervention $\bar{d}_t^2$) for a baseline sample size of 200; between 116 (intervention $\bar{d}_t^4$) and 238 (intervention $\bar{d}_t^2$) for a baseline sample size of 600; and between 190 (intervention $\bar{d}_t^4$) and 396 (intervention $\bar{d}_t^2$) for a baseline sample size of 1000.

After 12 months of follow-up, for all interventions super learning (for learner 3) made substantial use of additive models when estimating the Q-model: the respective super learner weight, averaged over time and simulation runs, was always greater than $0.73$. The g-models made particular use of GLM's selected by AIC.

\begin{figure}[ht!]
\begin{center}
\includegraphics[page=2,scale=0.625]{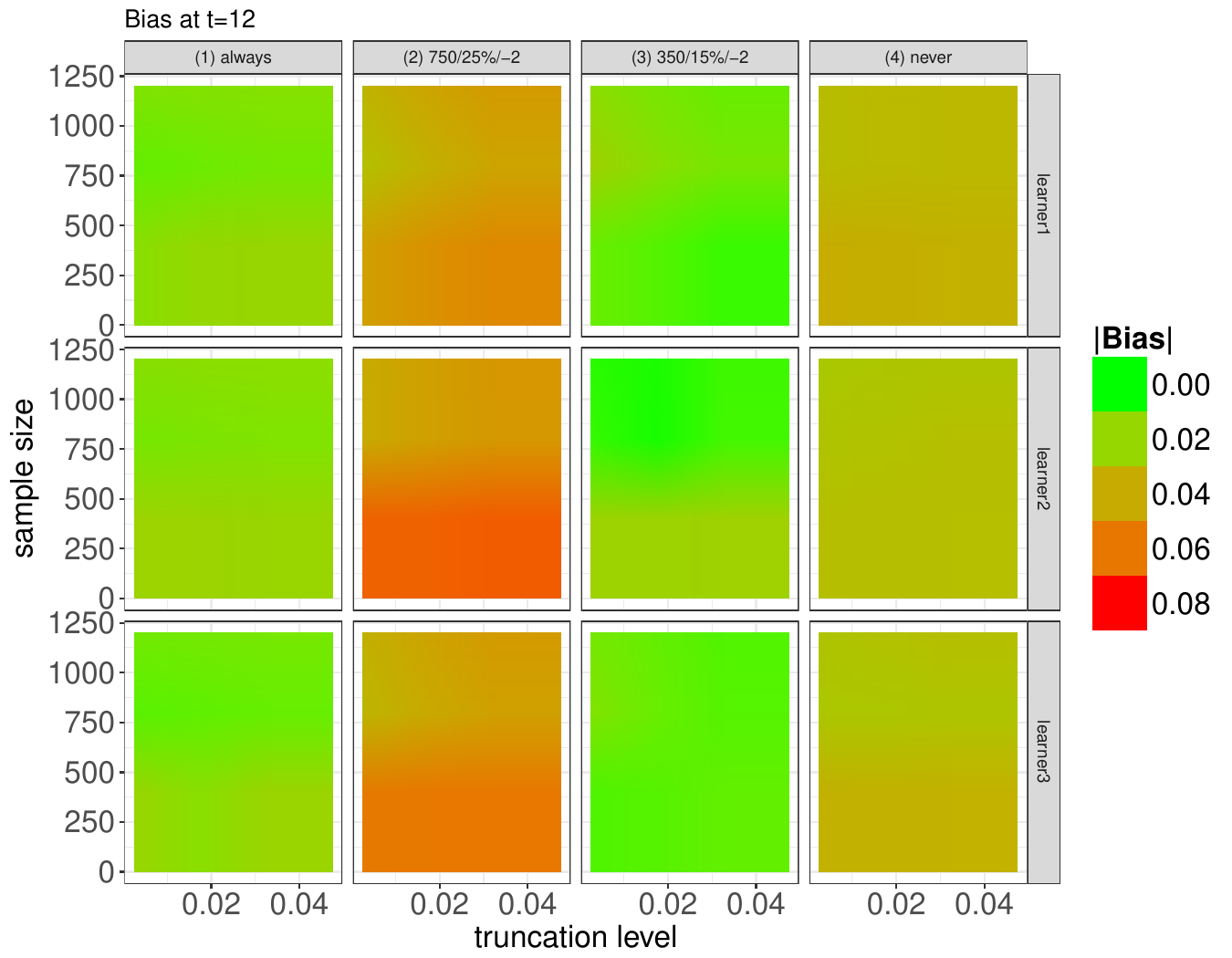}
\includegraphics[page=4,scale=0.625]{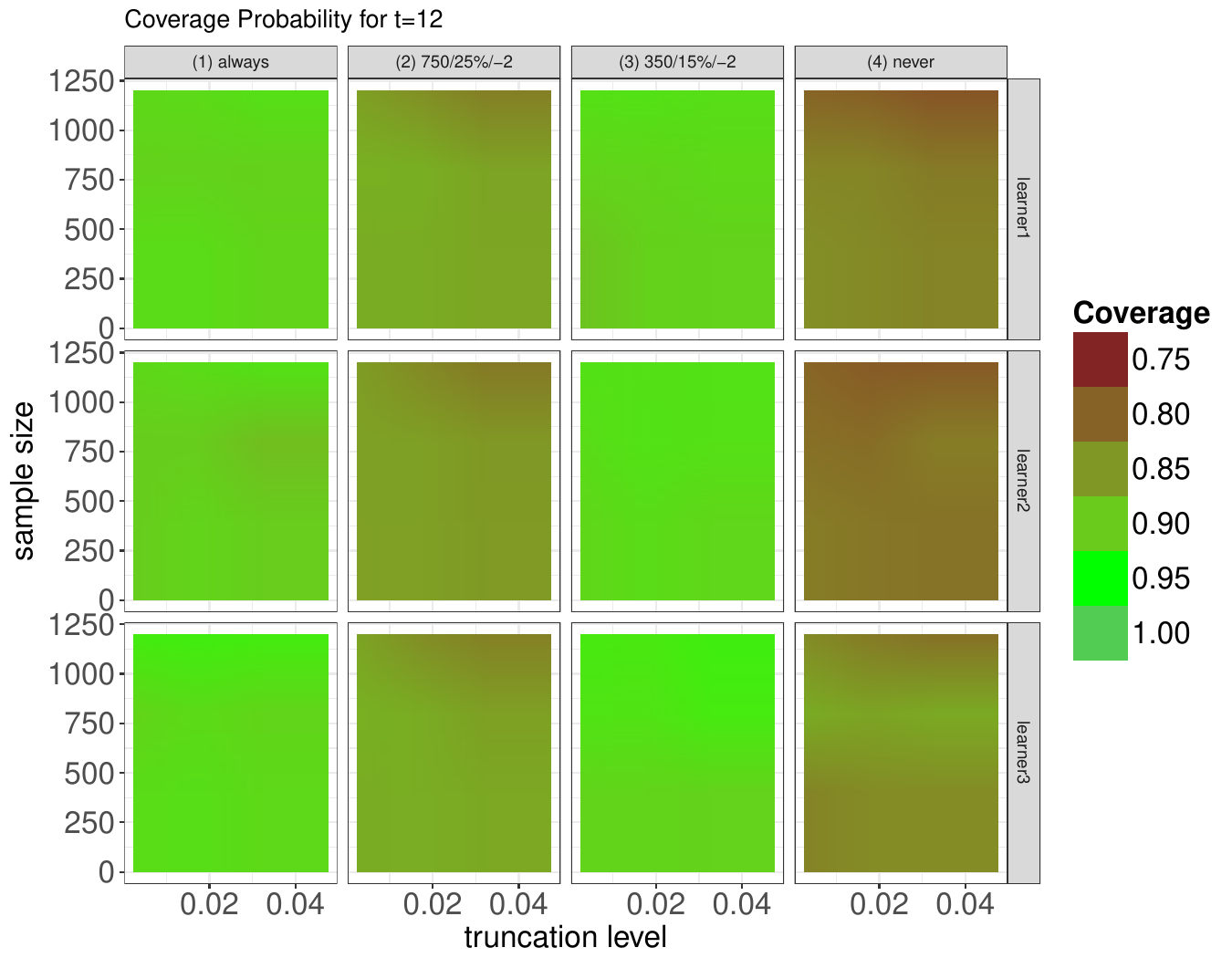}
\caption{Estimated bias and coverage probability, at $t=12$, for different truncation levels, interventions, learner sets, and sample sizes.} \label{figure:simulation}
\end{center}
\end{figure}

Figure \ref{figure:simulation} shows the estimated bias and coverage probability, for different truncation levels, interventions, learner sets, and sample sizes.

One can see that the bias varies substantially with respect to the chosen intervention. Interestingly the interventions with the biggest usable sample size and with the lowest usable sample size, $\bar{d}_t^4$ and $\bar{d}_t^2$, have the highest bias, showing that neither sample size nor type of intervention (static/dynamic) offers a reasonable guideline to judge the quality of the estimated target quantity. Interventions with high bias typically also have a poorer coverage. However, somewhat unexpectedly, for increasing sample size, coverage often decreases for the second and fourth intervention.

In this particular simulation setup the level of truncation hardly affects the results. There are gains with a larger sample size, as one would expect; it is however worth pointing out that even with small sample sizes results are relatively stable and good.

Of course, learner 3, with the biggest set of learners, typically performs best. Surprisingly the other two learners also yield good results -- which is encouraging given that the data-generating process contains interactions and non-linear associations as well, and these effects are not specifically modeled by learner set 1 and learner set 2.

Since in the simulation the data-generating process is known, we can estimate $P(A_t=\bar{d}_t|\bar{\mathbf{L}}_t=\bar{\mathbf{l}}_t,\bar{A}_{t-1}=\bar{d}_{t-1})$, i.e. the probability of continuing to receive treatment according to the assigned treatment rule, given that a patient has received treatment so far and irrespective of the covariate history. That is, we can look at the indicator variable whether a patient has been treated according to the rule of interest at each time point $t$, or not, given the past covariates and conditional on those who were treated according to the rule at $t-1$. This approach measures the ``data support'' for a given rule of interest. If a substantial proportion of subjects have very small cumulative probabilities, then this could be a sign of positivity violations. Table \ref{tab:positivity} lists the summaries of the cumulative probabilities.

\begin{table}[ht!]
\caption{Proportion of cumulative probabilities of continuing to receive treatment according to the assigned treatment rule (given that this has been done so far and irrespective of the covariate history) which are smaller than $0.025$. The numbers are calculated based on the data-generating process in the simulation, and are estimated from the data set as well, and for both at the end of follow-up.}
\label{tab:positivity}
\begin{center}
\begin{tabular}{r|cccc}
& $\bar{d}^1$ & $\bar{d}^2$ & $\bar{d}^3$ & $\bar{d}^4$\\
\hline
simulation & 0.3\% & 1.0\% & 0.6\% & 1.5\% \\
data       & 0\%   & 2.1\% & 0.5\% & 3.0\% \\
\end{tabular}
\end{center}
\end{table}

The table clearly shows that the highest proportion of small cumulative probabilities is found for the second and fourth intervention, i.e. those interventions which performed worst in the simulations. This highlights the fact that interventions need data support to meet the practical positivity assumption. In general this can also be estimated in a data analysis, for a given estimation technique. When using a standard logistic regression model we obtain the highest proportion of small cumulative probabilities for the fourth intervention, i.e. the intervention that was most volatile in the data analysis. We suggest that in addition to summaries of the clever covariates and percentage of truncated values, the data support for each intervention, as presented in Table \ref{tab:positivity}, should also serve as  a diagnostic tool in LTMLE analyses.

\section{Conclusions}\label{sec:conclusion}
To our knowledge this is one of the first successful implementations of LTMLE for a data set with $>10$ follow-up times \textit{and} at least $3$ time-dependent confounders (which are affected by prior treatment). Moreover, our data analysis was based on a realistic data set relevant to public health policy making, and was the first comparison between the (parametric) g-formula and LTMLE in such a complex context. Large longitudinal data sets require specific care when using LTMLE: we have shown that long follow-up may come with sample sizes that are small compared to the number of parameters that need to be fitted for machine learning algorithms, including semi-parametric regression models. This can yield instabilities, memory problems and identifiability problems that need to be addressed.

We have argued that there is practically no way around model selection in causal inference with both the g-formula and LTMLE, despite the implicit assumption that some of the included confounders from the structural causal model are irrelevant. This is because the causal parameter of interest can be best identified with good predictions coming from an optimal bias-variance tradeoff (though super learning provides a good bias-variance tradeoff for the nuisance models and not necessarily the target parameter and estimation approaches such as collaborative TMLE \cite{Schnitzer:2017} have been suggested). This model selection process could incorporate clinical knowledge. We have included non-linear interactions using additive models to address such hypotheses. While this may have potential benefits for the g-formula, our LTMLE estimator could not be improved by such a strategy. In fact, a small super learner, with few algorithms, led to more reliable and precise results.

Ideally large longitudinal causal analyses are best analyzed by several methods. If the results match, then this is perfect. If not, diagnostics such as clever covariate and truncation summaries for LTMLE, or the natural course scenario for the g-formula are indispensable. Our analyses have shown that neither a good natural course scenario for the g-formula nor considerations of sample size and type of intervention (static/dynamic) offers a reasonable guideline to judge the quality of the estimated target quantity; what matters is the support in the data for the intervention rule of interest: high levels of truncation and small cumulative probabilities of continuing to follow the rule of interest are typically a sign of problems, likely due to practical positivity violations or poor modeling strategies. They can be addressed by an increase of appropriate learners, even if they are simple, or -- less preferably -- a change in the target quantity. Whatever the methodological choices are, it remains important to consider the substantive question at hand as well; for example, in our analysis, the most conservative estimates are provided by the g-formula, and a clinician may prefer to base her/his decisions on these estimates if the diagnostics for the method are acceptable.

We conclude the following with respect to the five questions posed in the introduction:
\begin{itemize}
\item it can be feasible to implement LTMLE in complex settings with long follow-up times, small sample size, multiple time-dependent confounders, and dynamic interventions.
\item as seen in both the data analysis and the simulations, LTMLE can be successful even when confronted with limited modeling options, for example by a limited number of applicable learners when using super learning. Nevertheless, this may be different in other study settings; and it would be beneficial to have more learners available that work with existing software packages and are computationally feasible in settings with long follow-up and therefore multiple variables. We have made a selection of such learners available \cite{Schomaker:2019}; and researchers with advanced programming capacities are advised to implement their own custom-made wrappers, tailored to their own analysis of interest, if possible.
\item in our setting, we couldn't find strong evidence that a flexible additive model, informed by prior clinical knowledge, possibly in conjunction with competing estimators such as the g-formula, may perform better than an automated LTMLE procedure with a limited set of learners.
\item in the settings studied, a reduced sample size or truncation of inverse probabilities didn't have a major impact on the results; but it was clear that different interventions may have different support in the data and that the success of estimators such as LTMLE also varied with respect to the chosen intervention. This highlights the need to develop more diagnostic tools to diagnose practical positivity violations in the data. Reporting the percentage of truncated cumulative inverse probabilities, summaries of the clever covariates and the data support for each intervention (see Section \ref{sec:simulation}) is highly recommended; but more research and guidance in this respect is needed.
\end{itemize}

\appendix

\section{$R$ code for automated estimation of $\Psi_6$ with \texttt{ltmle}}\label{sec:appendix_ltmle_aut}
To use the \texttt{ltmle} function from the package \texttt{ltmle}, one needs to provide the data in wide format. Below we show 5 patients of our data set, where $b1$, $b2$, $b3$, $b4$ refer to the baseline values of $\sqrt{\text{CD4 count}}$, CD4\%, WAZ, and HAZ.  Variable names ending with ``.$t^{\ast}$'' refer to measurements at time $t^{\ast}$ (with $t^{\ast}=0,1,3,6,9,\ldots,30$). A.0 and C.0 are not needed and reported as by definition everyone is ART-naive and uncensored at baseline. The choice of the data structure in terms of the ordering of the variables is discussed in Section \ref{sec:MvM_structure}.  \\

\begin{mdframed}[roundcorner=10pt, backgroundcolor=black!10, linecolor=black!10]
\footnotesize{
\begin{verbatim}
   gender   age_fv region2       b1 b2    b3    b4 cd4a_cf.1 cd4p_cf.1 wfa_cf.1 hfa_cf.1
1       1 4.752909       2 23.87467 15 -0.37 -0.30  23.87467        15    -0.37    -0.30
2       1 2.450377       2 26.98148 14 -0.81 -2.52  26.98148        14    -0.81    -2.52
3       1 2.318960       2 35.66511 14 -2.19 -3.13  35.66511        14    -2.19    -3.13
4       1 2.275154       2 34.46738 13 -2.13 -2.42  34.46738        13    -2.29    -2.38
5       1 3.526352       2 19.79899 26 -3.64 -2.52  19.79899        26    -3.66    -2.57

   ART.1 censored.1 cd4a_cf.3 cd4p_cf.3 wfa_cf.3 hfa_cf.3 ART.3 censored.3 cd4a_cf.6
1      0 uncensored  23.87467        15    -0.50    -0.53     1 uncensored  23.87467
2      0 uncensored  26.98148        14    -1.30    -2.20     1 uncensored  26.98148
3      0 uncensored  35.66511        14    -2.54    -3.08     1 uncensored  35.66511
4      1 uncensored  34.46738        13    -2.41    -2.28     1 uncensored  36.63332
5      0   censored        NA        NA       NA       NA    NA       <NA>        NA

   cd4p_cf.6 wfa_cf.6 hfa_cf.6
1         15    -0.50    -0.53
2         14    -1.36    -1.99
3         14    -2.72    -3.36
4         17    -2.41    -2.28
5         NA       NA       NA
\end{verbatim}
}
\end{mdframed}\vspace*{0.5cm}

Before applying the \texttt{ltmle} command we need to define the intervention matrix. For a static intervention, this could just be a vector such as $c(0,0)$, but needs to be a matrix for dynamic interventions as shown below. For illustrative purposes we use the intervention $\bar{d}_t^2$: ``starting ART if either CD4 count $<750$ cells/microlitre or CD4\% $<25$\%'' below. We also need to define the libraries used for the super learner, see Section \ref{sec:MvM_machine_learning} for more details. \\

\begin{mdframed}[roundcorner=10pt, backgroundcolor=black!10, linecolor=black!10]
\footnotesize{
\begin{verbatim}
# Define Intervention: start ART if CD4 count < 750 or CD4% < 25%
cd4_750.1  <- (mydata_wide$cd4a_cf.1 < sqrt(750)  | mydata_wide$cd4p_cf.1 < 25)
cd4_750.3  <- (mydata_wide$cd4a_cf.3 < sqrt(750)  | mydata_wide$cd4p_cf.3 < 25)   | cd4_750.1
abar750 <- as.matrix(cbind(cd4_750.1,cd4_750.3))
abar750[is.na(abar750)] <- 0

# Define collection of estimation methods to be used by SuperLearner
mylibrary <- list(Q=c("SL.mean","SL.glm","SL.stepAIC","SL.step.interaction","SL.gam","SL.bayesglm"),
                  g=c("SL.mean","SL.glm","SL.stepAIC","SL.gam","SL.bayesglm"))
\end{verbatim}
}
\end{mdframed}

Now, $\Psi_6$ can be estimated easily as follows:  \\

\begin{mdframed}[roundcorner=10pt, backgroundcolor=black!10, linecolor=black!10]
\footnotesize{
\begin{verbatim}
ltmle(mydata_wide[,1:23],
   Anodes=c("ART.1","ART.3"),
   Cnodes=c("censored.1","censored.3"),
   Lnodes=c("cd4a_cf.1","cd4p_cf.1","wfa_cf.1","cd4a_cf.3","cd4p_cf.3","wfa_cf.3",
            "cd4a_cf.6","cd4p_cf.6","wfa_cf.6"),
   Ynodes=c("hfa_cf.1","hfa_cf.3","hfa_cf.6"), Yrange=c(-15,15),
   Qform=NULL, gform=NULL, deterministic.g.function=MaintainTreatment, abar=abar750,
   SL.library=mylibrary, estimate.time=T, variance.method="ic", gbounds = c(0.01, 1)))
\end{verbatim}
}
\end{mdframed}

Note that \texttt{Anodes, Lnodes, Cnodes, Ynodes} refer to the treatment, censoring, covariate, and outcome data respectively. By specifying a \texttt{SL.library}, and leaving the arguments for model formulae (\texttt{Qform, gform}) empty, model specification is done by super learning. \texttt{Yrange=c(-15,15)} ensures that the outcome is bounded (i.e. absolute z-scores greater than 15 are regarded to be impossible).

\section{$R$ code for manual LTMLE estimation of $\Psi_6$ }\label{sec:appendix_ltmle_man}

The below implementation has been designed to be able to exactly reproduce the steps of the \texttt{ltmle} command; but to adhere to the framework and notation of Section \ref{sec:framework} at the same time. As noted in the manuscript, we work with a transformed outcome and quasi-binomial models as \texttt{ltmle} does. We also add the clever covariate as weights, as \texttt{ltmle} does. Furthermore, we use additive models based on the package \texttt{mgcv} for model fitting. For comparing the manual and automated implementation, one could simply use simple GLMs (no super learning) and specify the model form in the \texttt{Qform} and \texttt{gform} arguments. The below code to truncate weights and to calculate confidence intervals is used later on in the main code. \\

\begin{mdframed}[roundcorner=10pt, backgroundcolor=black!10, linecolor=black!10]
\footnotesize{
\begin{verbatim}
gbounds <- function(x,bounds=c(0.05,1)){
  x[x<min(bounds)] <- min(bounds)
  x[x>max(bounds)] <- max(bounds)
  return(x) }

CI <- function(est,infcurv){
n <- length(infcurv)
ciwidth <- 1.959964 * sd(infcurv)/sqrt(n)
CI <- c(est-ciwidth, est+ciwidth)
return(CI)  }
library(mgcv)
\end{verbatim}
}
\end{mdframed}

To estimate $\psi_6$ we need to do the following:\\

\begin{mdframed}[roundcorner=10pt, backgroundcolor=black!10, linecolor=black!10]
\footnotesize{
\begin{verbatim}
mydata_wide6 <-  mydata_wide[,1:23]
attach(mydata_wide6)

# Already prepare step 3c: fit treatment and censoring models
# (Note that i) the data after censoring are defined as missing by us and thus the censoring models are
# implicitly fit on the subset of those that are uncensored until the respective previous time point and
# ii) the models contain only variables that have been selected by cross validation [see Section 4.3.1])
m1_c   <- mgcv::gam(I(censored.1=="censored")  ~ age_fv  + s(b2) + s(b3) + s(b4), family=binomial)
m3_c   <- mgcv::gam(I(censored.3=="censored")  ~ gender + s(age_fv) + wfa_cf.1
                    + s(I((hfa_cf.1  +15)/30)), family=quasibinomial)

m1_a   <- mgcv::gam(ART.1 ~ s(cd4a_cf.1) + s(cd4p_cf.1) + s(wfa_cf.1) + s(b1) + s(b2) + s(b3),
                              data=mydata_wide6, family=binomial)
m3_a   <- mgcv::gam(ART.3 ~ s(cd4p_cf.3) + s(wfa_cf.3) + s(cd4a_cf.1) + s(cd4p_cf.1)  + s(b1) +
                              s(b2) + s(b3) + age_fv,
                              data=mydata_wide6[censored.1==1,], family=quasibinomial)

# Already prepare step 3c: define subsets of patients uncensored and following the treatment rule
# (Note that these subsets are needed to calculate the clever covariate as defined in (4))
subset1  <- censored.1=="uncensored" & ART.1==abar[,1]
subset3  <- censored.1=="uncensored" & ART.1==abar[,1] & censored.3=="uncensored" & ART.3==abar[,2]

# Still prepare data for step 3c: calculate the cumulative treatment and censoring probabilites;
#  also: apply cutoff at 0.05, i.e. probabilities <0.05 are defined as 0.05 -> clever covariate <= 20
p1a <- gbounds((ART.1 == 1)*predict(m1_a, type="response", newdata=mydata_wide6) +
               (ART.1 == 0)*(1 - predict(m1_a, type="response", newdata=mydata_wide6)))
p3a <- gbounds((ART.3 == 1)*predict(m3_a, type="response", newdata=mydata_wide6) +
               (ART.3 == 0)*(1 - predict(m3_a, type="response", newdata=mydata_wide6)))

p1c <- gbounds((censored.1 == "uncensored")*(1 - predict(m1_c, type="response",newdata=mydata_wide6)))
p3c <- gbounds((censored.3 == "uncensored")*(1 - predict(m3_c, type="response",newdata=mydata_wide6)))

# Still prepare data for step 3c: calculate clever covariate (=0 [weight 0] for those excluded)
# (i.e. the inverse probabilities are only used for the subset of those patients
# following the treatment rule and are uncensored, see equation (4))
mydata_wide6$cc1 <- mydata_wide6$cc3 <- 0
mydata_wide6$cc1[subset1] <- (1/(gbounds(p1a*p1c)))[subset1]
mydata_wide6$cc3[subset3] <- (1/(gbounds(p1a*p1c*p3a*p3c)))[subset3]

# Prepare for step 2: Define Intervention
newdata_wide6 <- mydata_wide6
newdata_wide6$ART.1 <- abar[,1]
newdata_wide6$ART.3 <- abar[,2]

### LTMLE estimation ####

# time point 6: start to calculate innermost expectation
m6_y <- mgcv::gam(I((hfa_cf.6 +15)/30)   ~ s(age_fv) + s(b2) + s(b3) + s(b4) + ART.1 + s(cd4p_cf.3)
                 + s(wfa_cf.3, by=b3) + s(I((hfa_cf.3 +15)/30)), data=mydata_wide6[censored.3==
                 "uncensored" & censored.1=="uncensored",], family=quasibinomial)  # STEP 1
mydata_wide$ov6 <- predict(m6_y, newdata=newdata_wide6,type="response")            # STEP 2
Qk6 <-  glm("I((hfa_cf.6 +15)/30)~  1 + offset(qlogis(ov6)) ", data=mydata_wide6, weights=cc3,
                                    family = quasibinomial)                        # STEP 3a-c
mydata_wide$y3.Q <- predict(Qk6, type="response",newdata=mydata_wide6)             # STEP 3d

# time point 3
m3_y <- mgcv::gam(y3.Q   ~ (age_fv) + s(b3) + s(b4) + ART.1 + s(wfa_cf.1,by=b3)
                  + s(I((hfa_cf.1 +15)/30),by=b3), data=mydata_wide6[censored.1=="uncensored",],
                   family=quasibinomial)                                          # STEP 1
mydata_wide$ov3 <- predict(m3_y, newdata=newdata_wide6,type="response")           # STEP 2
Qk3 <-  glm("y3.Q ~  1 + offset(qlogis(ov3)) ", data=mydata_wide6, weights=cc1,   # STEP 3a-c
            family = quasibinomial)
mydata_wide$y1.Q <- predict(Qk3, type="response",newdata=mydata_wide6)            # STEP 3d

# time point 1+0: calculate final TMLE estimate
# (as everything before ART.1/hfa_cf.1 is essentially baseline due to no prior intervention)
est_ltmle.6 <- mean(mydata_wide$y1.Q)*30-15                                 # STEP 4

# Confidence Interval: calculate IC and set HAZ=0 if unmeasured at time t   # Step 5
mydata_wide6$Y <- (mydata_wide6$hfa_cf.6+15)/30
mydata_wide6$Y[is.na(mydata_wide6$Y)]<-0
mydata_wide6$y3.Q[is.na(mydata_wide6$y3.Q)]<-0
IC <- mydata_wide6$cc3*(mydata_wide6$Y-mydata_wide6$y3.Q) + mydata_wide6$cc1*
     (mydata_wide6$y3.Q-mydata_wide6$y1.Q)  + mydata_wide6$y1.Q-mean(mydata_wide6$y1.Q)
IC <- IC*30-15
CI(est_ltmle.6,IC)  # Final Confidence Interval
\end{verbatim}
}
\end{mdframed}

\section{Data generating process in the simulation study}\label{sec:appendix_data_generating}
Both baseline data ($t=0$) and follow-up data ($t=1,\ldots,12$) were created using structural equations using the $R$-package \texttt{simcausal} \cite{Sofrygin:2016}. The below listed distributions, listed in temporal order, describe the data-generating process. Baseline data $\mathbf{L}_0$ refers to region, sex, age, CD4 count, CD4\%, WAZ and HAZ respectively ($V^1$, $V^2$, $V^3$, $L^1_0$, $L^2_0$, $L^3_0$, $Y_0$). Follow-up data refers to CD4 count, CD4\%, WAZ and HAZ ($L^1_t$, $L^2_t$, $L^3_t$, $Y_t$), as well as a treatment ($A_t$) and censoring ($C_t$) indicator. In addition to Bernoulli ($B$), uniform ($U$) and normal ($N$) distributions, we also use truncated normal distributions which are denoted by $N_{[a,b]}$ where $a$ and $b$ are the truncation levels. Values which are smaller than $a$ are replaced by a random draw from a $U(a_1,a_2)$ distribution and values greater than $b$ are drawn from a $U(b_1,b_2)$ distribution. Values for $(a_1, a_2, b_1, b_2)$ are $(0,50,5000,10000)$ for $L^1$, (0.03,0.09,0.7,0.8) for $L^2$, and $(-10,3,3,10)$ for both $L^3$ and $Y$. The notation $\bar{\mathcal{D}}$ means ``conditional on the data that has already been measured (generated) according to the time ordering''.\\

{\footnotesize{
For $t=0$:
\begin{eqnarray*}
V^1 &\sim& B(p=4392/5826) \\
V^2|\bar{\mathcal{D}} &\sim& \left\{ \begin{array}{cl}
               B(p=2222/4392) & \quad \mbox{if} \quad V^1 = 1\\
               B(p=758/1434)  & \quad \mbox{if} \quad V^1 = 0\\
               \end{array}
               \right. \\
V^3|\bar{\mathcal{D}} &\sim& U(1,5) \\
L^1_0|\bar{\mathcal{D}} &\sim & \left\{ \begin{array}{cl}
               N_{[0,10000]}(650,350) & \quad \mbox{if} \quad V^1 = 1\\
               N_{[0,10000]}(720,400))  & \quad \mbox{if} \quad V^1 = 0\\
               \end{array}
               \right. \\
\tilde{L}^1_0|\bar{\mathcal{D}} &\sim & N((L^1_0-671.7468)/(10\cdot352.2788)+1,0)\\
L^2_0|\bar{\mathcal{D}} &\sim &  N_{[0.06,0.8]}(0.16+0.05\cdot(L^1_0-650)/650,0.07) \\
\tilde{L}^2_0|\bar{\mathcal{D}} &\sim & N((L^2_0-0.1648594)/(10\cdot0.06980332)+1,0)\\
L^3_0|\bar{\mathcal{D}} &\sim & \left\{ \begin{array}{cl}
               N_{[-5,5]}(- 1.65 + 0.1 \cdot V^3 + 0.05 \cdot (L^1_0 - 650)/650 + 0.05 \cdot (L^2_0 - 16)/16,1) & \quad \mbox{if} \quad V^1 = 1\\
               N_{[-5,5]}( -2.05 + 0.1 \cdot V^3 + 0.05 \cdot (L^1_0 - 650)/650 + 0.05 \cdot (L^2_0 - 16)/16,1))  & \quad \mbox{if} \quad V^1 = 0\\
               \end{array}
               \right. \\
A_0|\bar{\mathcal{D}} &\sim& B(p=0) \\
C_0|\bar{\mathcal{D}} &\sim& B(p=0) \\
Y_0|\bar{\mathcal{D}} &\sim &  N_{[-5,5]}(-2.6 + 0.1 \cdot I(V^3 > 2) + 0.3 \cdot I(V^1 = 0) + (L^3_0 + 1.45),1.1) \\
\end{eqnarray*}

\clearpage
For $t>0$:

\begin{eqnarray*}
L^1_t|\bar{\mathcal{D}} &\sim & \left\{ \begin{array}{cl}
               N_{[0,10000]}(13\cdot\log(t \cdot (1034-662)/8) + L^1_{t-1} + 2 \cdot L^2_{t-1} + 2 \cdot L^3_{t-1} + 2.5 \cdot A_{t-1},50) & \quad \mbox{if} \quad t \in \{1,2,3,4\}\\
               N_{[0,10000]}(4\cdot\log(t \cdot (1034-662)/8) + L^1_{t-1} + 2 \cdot L^2_{t-1} + 2 \cdot L^3_{t-1} + 2.5 \cdot A_{t-1},50) & \quad \mbox{if} \quad t \in \{5,6,7,8\}\\
               N_{[0,10000]}(L^1_{t-1} + 2 \cdot L^2_{t-1} + 2 \cdot L^3_{t-1} + 2.5 \cdot A_{t-1},50) & \quad \mbox{if} \quad t \in \{9,10,11,12\}\\
               \end{array}
               \right. \\
L^2_t|\bar{\mathcal{D}} &\sim &  N_{[0.06,0.8]}(L^2_{t-1} + 0.0003 \cdot (L^1_t - L^1_{t-1}) + 0.0005 \cdot (L^3_{t-1}) + 0.0005 \cdot A_{t-1} \cdot \tilde{L}^1_0,0.02) \\
L^3_t|\bar{\mathcal{D}} &\sim &  N_{-5,5}(L^3_{t-1} + 0.0017 \cdot (L^1_t - L^1_{t-1}) + 0.2 \cdot (L^2_t - L^2_{t-1}) + 0.005 \cdot A_{t-1} \cdot \tilde{L}^2_0,0.5) \\
A_t|\bar{\mathcal{D}}  &\sim& \left\{ \begin{array}{cl}
               B(p=1) & \quad \mbox{if} \quad A_{t-1} = 1\\
               B(p=1/(1+\exp(-[-2.4 + 0.015 \cdot (750 - L^1_t) + 5 \cdot (0.2 - L^2_t) - 0.8 \cdot L^3_t + 0.8 \cdot t])))  & \quad \mbox{if} \quad A_{t-1} = 0\\
               \end{array}
               \right.  \\
C_t|\bar{\mathcal{D}} &\sim& B(p=1/(1+\exp(-[-6+ 0.01 \cdot (750 - L^1_t) + 1 \cdot (0.2 - L^2_t) - 0.65 \cdot L^3_t - A_t]))) \\
Y_t|\bar{\mathcal{D}} &\sim &  N_{[-5,5]}(Y_{t-1} +
         0.00005 \cdot (L^1_t - L^1_{t-1}) - 0.000001 \cdot \left((L^1_t  - L^1_{t-1})\cdot \sqrt{\tilde{L}^1_0}\right)^2 +
         0.01 \cdot (L^2_t - L^2_{t-1})- \\
         &&  0.0001 \cdot \left((L^2_t - L^2_{t-1})\cdot \sqrt{\tilde{L}^2_0}\right)^2 + 0.07 \cdot ((L^3_t-L^3_{t-1})\cdot(L^3_0+1.5135)) - 0.001 \cdot ((L^3_t-L^3_{t-1})\cdot(L^3_0+1.5135))^2 + \\
         && 0.005 \cdot A_t + 0.075 \cdot A_{t-1} + 0.05 \cdot A[t] \cdot A[t-1] ,0.01) \\
\end{eqnarray*}
}}

\section*{Acknowledgements}
We are immensely grateful to Maya Petersen for her constructive and helpful feedback on earlier versions of this manuscript. We appreciate the continuous help of Joshua Schwab who helped solving multiple issues related to the implementation of LTMLE, particularly in conjunction with the $R$ package \texttt{ltmle}. We further thank Matthias A{\ss}enmacher for his dedicated help in constructing the simulation setup. We also thank Lorna Renner, Shobna Sawry, Sylvie N'Gbeche, Karl-G{\"u}nter Technau, Francois Eboua, Frank Tanser, Haby Syg\-nate-Sy, Sam Phiri, Madeleine Amorissani-Folquet, Vivian Cox, Fla Koueta, Cleophas Chimbete, Annette Lawson-Evi, Janet Giddy, Clarisse Amani-Bosse, and Robin Wood for sharing their data with us. Computations were performed using facilities provided by the University of Cape Town's ICTS High Performance Computing team.  We would also like to highlight the support of the International epidemiologic Databases to Evaluate AIDS Southern Africa (IeDEA-SA) and the West Africa Paediatric Working Group. The NIH has supported the above individuals, grant numbers U01AI069924 and U01AI069919. Miguel Angel Luque Fernandez is supported by the Spanish National Institute of Health, Carlos III Miguel Servet I Investigator Award (CP17/00206).

\clearpage
%%%%%%%%%%%%%%%%%%
\bibliographystyle{unsrt}
{\footnotesize
\bibliography{LTMLE}

\begin{thebibliography}{10}

\bibitem{Gruber:2010b}
S.~Gruber and M.~J. van~der Laan.
\newblock An application of collaborative targeted maximum likelihood
  estimation in causal inference and genomics.
\newblock {\em Int J Biostat}, 6(1):Article 18, 2010.

\bibitem{Pearl:2016}
M.~Pearl, L.~Balzer, and J.~Ahern.
\newblock Targeted estimation of marginal absolute and relative associations in
  case-control data: An application in social epidemiology.
\newblock {\em Epidemiology}, 27(4):512--517, 2016.

\bibitem{Luque:2017}
M.~A. Luque-Fernandez, A.~Belot, L.~Valeri, G.~Ceruli, C.~Maringe, and
  B.~Rachet.
\newblock Data-adaptive estimation for double-robust methods in
  population-based cancer epidemiology: Risk differences for lung cancer
  mortality by emergency presentation.
\newblock {\em American Journal of Epidemiology}, 187:871--878, 2017.

\bibitem{Luque:2018}
M.~A. Luque-Fernandez, M.~Schomaker, B.~Rachet, and M.~Schnitzer.
\newblock Targeted maximum likelihood estimation for a binary treatment: A
  tutorial.
\newblock {\em Statistics in Medicine}, 37:2530--2546, 2018.

\bibitem{Schomaker:2018}
C.~Gehringer, H.~Rode, and M.~Schomaker.
\newblock The effect of electrical load shedding on pediatric hospital
  admissions in {S}outh {A}frica.
\newblock {\em Epidemiology}, 29:841--847, 2018.

\bibitem{Decker:2014}
A.~Decker, A.~Hubbard, C.~Crespi, E.~Seto, and M~Wang.
\newblock Semiparametric estimation of the impacts of longitudinal
  interventions on adolescent obesity using targeted maximum-likelihood:
  Accessible estimation with the ltmle package.
\newblock {\em Journal of Causal Inference}, 2(1):95--108, 2014.

\bibitem{Gianfrancesco:2016}
M.~A. Gianfrancesco, L.~Balzer, K.~E. Taylor, L.~Trupin, J.~Nititham, M.~F.
  Seldin, A.~W. Singer, L.~A. Criswell, and L.~F. Barcellos.
\newblock Genetic risk and longitudinal disease activity in systemic lupus
  erythematosus using targeted maximum likelihood estimation.
\newblock {\em Genes and Immunity}, 17(6):358--62, 2016.

\bibitem{Hubbard:2012}
A.~Hubbard, F.~Jamshidian, and N.~Jewell.
\newblock Adjusting for perception and unmasking effects in longitudinal
  clinical trials.
\newblock {\em International Journal of Biostatistics}, 8(2):7, 2012.

\bibitem{Kreif:2017}
N.~Kreif, L.~Tran, R.~Grieve, B.~deStavola, R.~Tasker, and M.~Petersen.
\newblock Estimating the comparative effectiveness of feeding interventions in
  the paediatric intensive care unit: a demonstration of longitudinal targeted
  maximum likelihood estimation.
\newblock {\em American Journal of Epidemiology}, 186:1370--1379, 2017.

\bibitem{Lendle:2017}
S.~Lendle, J.~Schwab, M.~Petersen, and M.~Van~der Laan.
\newblock ltmle: An {R} package implementing targeted minimum loss-based
  estimation for longitudinal data.
\newblock {\em Journal of Statistical Software}, 81:1--21, 2017.

\bibitem{Neugebauer:2014}
R.~Neugebauer, J.~A. Schmittdiel, and M.~J. van~der Laan.
\newblock Targeted learning in real-world comparative effectiveness research
  with time-varying interventions.
\newblock {\em Statistics in Medicine}, 33(14):2480--520, 2014.

\bibitem{Petersen:2014}
M.~Petersen, J.~Schwab, S.~Gruber, N.~Blaser, M.~Schomaker, and M.~van~der
  Laan.
\newblock Targeted maximum likelihood estimation for dynamic and static
  longitudinal marginal structural working models.
\newblock {\em Journal of Causal Inference}, 2:147--185, 2014.

\bibitem{Petersen:2014a}
M.~L. Petersen.
\newblock Commentary: Applying a causal road map in settings with
  time-dependent confounding.
\newblock {\em Epidemiology}, 25(6):898--901, 2014.

\bibitem{Schnitzer:2014}
M.~E. Schnitzer, E.~E. Moodie, M.~J. van~der Laan, R.~W. Platt, and M.~B.
  Klein.
\newblock Modeling the impact of hepatitis {C} viral clearance on end-stage
  liver disease in an {HIV} co-infected cohort with targeted maximum likelihood
  estimation.
\newblock {\em Biometrics}, 70(1):144--52, 2014.

\bibitem{Schnitzer:2014b}
M.~E. Schnitzer, M.~J. van~der Laan, E.~E. Moodie, and R.~W. Platt.
\newblock Effect of breastfeeding on gastrointestinal infection in infants: A
  targeted maximum likelihood approach for clustered longitudinal data.
\newblock {\em Annals of Applied Statistics}, 8(2):703--725, 2014.

\bibitem{Schnitzer:2016}
Mireille~E. Schnitzer, Judith~J. Lok, and Ronald~J. Bosch.
\newblock Double robust and efficient estimation of a prognostic model for
  events in the presence of dependent censoring.
\newblock {\em Biostatistics}, 17(1):165--177, 2016.

\bibitem{Stitelman:2012}
O.~M. Stitelman, V.~De~Gruttola, and M.~J. van~der Laan.
\newblock A general implementation of tmle for longitudinal data applied to
  causal inference in survival analysis.
\newblock {\em International Journal of Biostatistics}, 8(1), 2012.

\bibitem{Tran:2016}
L.~Tran, C.~Yiannoutsos, B.~Musick, K.~Wools-Kaloustian, A.~Siika, S.~Kimaiyo,
  M.~Van~der Laan, and M.~L. Petersen.
\newblock Evaluating the impact of a {HIV} low-risk express care task-shifting
  program: A case study of the targeted learning roadmap.
\newblock {\em Epidemiological Methods}, 5(1):69--91, 2016.

\bibitem{Tran:2019}
L.~Tran, C.~Yiannoutsos, K.~Wools-Kaloustian, A.~Siika, M.~van~der Laan, and
  M.~Petersen.
\newblock Double robust efficient estimators of longitudinal treatment effects:
  Comparative performance in simulations and a case study.
\newblock {\em International Journal of Biostatistics}, in press, 2019.

\bibitem{Robins:2000}
J.~M. Robins, M.~A. Hernan, and B.~Brumback.
\newblock Marginal structural models and causal inference in epidemiology.
\newblock {\em Epidemiology}, 11(5):550--560, 2000.

\bibitem{Bang:2005}
H.~Bang and J.~M. Robins.
\newblock Doubly robust estimation in missing data and causal inference models.
\newblock {\em Biometrics}, 64(2):962--972, 2005.

\bibitem{Robins:1986}
J.~Robins.
\newblock A new approach to causal inference in mortality studies with a
  sustained exposure period - application to control of the healthy worker
  survivor effect.
\newblock {\em Mathematical Modelling}, 7(9-12):1393--1512, 1986.

\bibitem{vanderLaan:2012}
M.~J. van~der Laan and S.~Gruber.
\newblock Targeted minimum loss based estimation of causal effects of multiple
  time point interventions.
\newblock {\em International Journal of Biostatistics}, 8(1):article 9, 2012.

\bibitem{vanderLaan:2011}
M.~Van~der Laan and S.~Rose.
\newblock {\em Targeted Learning}.
\newblock Springer, 2011.

\bibitem{Stefanski:2002}
L.~A. Stefanski and D.~D. Boos.
\newblock The calculus of m-estimation.
\newblock {\em American Statistician}, 56(1):29--38, 2002.

\bibitem{Edmonds:2011}
A.~Edmonds, M.~Yotebieng, J.~Lusiama, Y.~Matumona, F.~Kitetele, S.~Napravnik,
  S.~R. Cole, A.~Van~Rie, and F.~Behets.
\newblock The effect of highly active antiretroviral therapy on the survival of
  {HIV-}infected children in a resource-deprived setting: a cohort study.
\newblock {\em PLoS Medicine}, 8(6):e1001044, 2011.

\bibitem{Violari:2008}
A.~Violari, M.~F. Cotton, D.~M. Gibb, A.~G. Babiker, J.~Steyn, S.~A. Madhi,
  P.~Jean-Philippe, and J.~A. McIntyre.
\newblock Early antiretroviral therapy and mortality among {HIV}-infected
  infants.
\newblock {\em New England Journal of Medicine}, 359(21):2233--2244, 2008.

\bibitem{Daniel:2013}
R.~M. Daniel, S.~N. Cousens, B.~L. De~Stavola, M.~G. Kenward, and J.~A. Sterne.
\newblock Methods for dealing with time-dependent confounding.
\newblock {\em Statistics in Medicine}, 32(9):1584--618, 2013.

\bibitem{Phutanakit:2012}
T.~Puthanakit, V.~Saphonn, J.~Ananworanich, P.~Kosalaraksa, R.~Hansudewechakul,
  U.~Vibol, S.~J. Kerr, S.~Kanjanavanit, C.~Ngampiyaskul, J.~Wongsawat,
  W.~Luesomboon, N.~Ngo-Giang-Huong, K.~Chettra, T.~Cheunyam, T.~Suwarnlerk,
  S.~Ubolyam, W.~T. Shearer, R.~Paul, L.~M. Mofenson, L.~Fox, M.~G. Law, D.~A.
  Cooper, P.~Phanuphak, M.~C. Vun, and K.~Ruxrungtham.
\newblock Early versus deferred antiretroviral therapy for children older than
  1 year infected with hiv (predict): a multicentre, randomised, open-label
  trial.
\newblock {\em Lancet Infectious Diseases}, 12(12):933--41, 2012.

\bibitem{Robins:2009}
J.~Robins and M.~A. Hernan.
\newblock Estimation of the causal effects of time-varying exposures.
\newblock In G.~Fitzmaurice, M.~Davidian, G.~Verbeke, and G.~Molenberghs,
  editors, {\em Longitudinal Data Analysis}, pages 553--599. CRC Press, 2009.

\bibitem{Daniel:2011}
R.M. Daniel, B.L. De~Stavola, and S.N. Cousens.
\newblock G- formula: Estimating causal effects in the presence of time-varying
  confounding or mediation using the g-computation formula.
\newblock {\em The Stata Journal}, 11(4):479--517, 2011.

\bibitem{Young:2011}
J.~G. Young, L.~E. Cain, J.~M. Robins, E.~J. O'Reilly, and M.~A. Hernan.
\newblock Comparative effectiveness of dynamic treatment regimes: an
  application of the parametric g-formula.
\newblock {\em Statistics in Biosciences}, 3(1):119--143, 2011.

\bibitem{Robins:2004}
J.~Robins, M.~A. Hernan, and U.~Siebert.
\newblock Effects of multiple interventions.
\newblock In M.~Ezzati, C.~Murray, and A.~Lopez, editors, {\em Comparative
  quantification of health risks: global and regional burden of disease
  attributable to selected major risk factors}, pages 2191--2230. World Health
  Organization, 2004.

\bibitem{Schomaker:2017}
M.~Schomaker, V.~Leroy, T.~Wolfs, K.~G. Technau, L.~Renner, A.~Judd, S.~Sawry,
  M.~Amorissani-Folquet, A.~Noguera-Julian, F.~Tanser, F.~Eboua, M.~L. Navarro,
  C.~Chimbetete, C.~Amani-Bosse, J.~Warszawski, S.~Phiri, S.~N'Gbeche, V.~Cox,
  F.~Koueta, J.~Giddy, H.~Sygnate-Sy, D.~Raben, G.~Chene, M.~A. Davies, D.~E.
  A.~West Ie, collaborations Southern Africa~regional, and Cohere~in EuroCoord.
\newblock Optimal timing of antiretroviral treatment initiation in hiv-positive
  children and adolescents: a multiregional analysis from {S}outhern {A}frica,
  {W}est {A}frica and {E}urope.
\newblock {\em International Journal of Epidemiology}, 46:453--465, 2017.

\bibitem{vanderLaan:2018}
M.~Van~der Laan and S.~Rose.
\newblock {\em Targeted Learning in Data Science: Causal Inference for Complex
  Longitudinal Studies}.
\newblock Springer, 2018.

\bibitem{Schnitzer:2017}
M.~E. Schnitzer and M.~Cefalu.
\newblock Collaborative targeted learning using regression shrinkage.
\newblock {\em Statistics in Medicine}, 37:530--543, 2017.

\bibitem{Breiman:1996}
L.~Breiman.
\newblock Stacked regressions.
\newblock {\em Machine Learning}, 24(1):49--64, 1996.

\bibitem{Polley:2017}
Eric Polley, Erin LeDell, Chris Kennedy, and Mark {van der Laan}.
\newblock {\em SuperLearner: Super Learner Prediction}, 2017.
\newblock {R} package version 2.0-22.

\bibitem{vanderLaan:2008}
M.~Van~der Laan, E.~Polley, and A.~Hubbard.
\newblock Super learner.
\newblock {\em Statistical Applications in Genetics and Molecular Biology},
  6:Article 25, 2008.

\bibitem{Egger:2012}
M.~Egger, D.~K. Ekouevi, C.~Williams, R.~E. Lyamuya, H.~Mukumbi, P.~Braitstein,
  T.~Hartwell, C.~Graber, B.~H. Chi, A.~Boulle, F.~Dabis, and
  K.~Wools-Kaloustian.
\newblock Cohort profile: The international epidemiological databases to
  evaluate aids ({IeDEA}) in sub-{S}aharan {A}frica.
\newblock {\em International Journal of Epidemiology}, 41(5):1256--1264, 2012.

\bibitem{Ekouevi:2011}
D.~K. Ekouevi, A.~Azondekon, F.~Dicko, K.~Malateste, P.~Toure, F.~T. Eboua,
  K.~Kouadio, L.~Renner, K.~Peterson, F.~Dabis, H.~S. Sy, and V.~Leroy.
\newblock 12-month mortality and loss-to-program in antiretroviral-treated
  children: The {IeDEA} pediatric {W}est {A}frican {D}atabase to evaluate
  {AIDS} ({pWADA}), 2000-2008.
\newblock {\em BMC Public Health}, 11:519, 2011.

\bibitem{Schomaker:2013}
M.~Schomaker, M.~Egger, J.~Ndirangu, S.~Phiri, H.~Moultrie, K.~Technau, V.~Cox,
  J.~Giddy, C.~Chimbetete, R.~Wood, T.~Gsponer, C.~Bolton~Moore, H.~Rabie,
  B.~Eley, L.~Muhe, M.~Penazzato, S.~Essajee, O.~Keiser, and M.~A. Davies.
\newblock When to start antiretroviral therapy in children aged 2-5 years: a
  collaborative causal modelling analysis of cohort studies from {S}outhern
  {A}frica.
\newblock {\em Plos Medicine}, 10(11):e1001555, 2013.

\bibitem{Schomaker:2016}
M.~Schomaker, M.~A. Davies, K.~Malateste, L.~Renner, S.~Sawry, S.~N'Gbeche,
  K.~Technau, F.~T. Eboua, F.~Tanser, H.~Sygnate-Sy, S.~Phiri,
  M.~Amorissani-Folquet, V.~Cox, F.~Koueta, C.~Chimbete, A.~Lawson-Evi,
  J.~Giddy, C.~Amani-Bosse, R.~Wood, M.~Egger, and V.~Leroy.
\newblock Growth and mortality outcomes for different antiretroviral therapy
  initiation criteria in children aged 1-5 years: A causal modelling analysis
  from {W}est and {S}outhern {A}frica.
\newblock {\em Epidemiology}, 237-246, 2016.

\bibitem{Puthanakit:2012}
T.~Puthanakit, V.~Saphonn, J.~Ananworanich, P.~Kosalaraksa, R.~Hansudewechakul,
  U.~Vibol, S.~J. Kerr, S.~Kanjanavanit, C.~Ngampiyaskul, J.~Wongsawat,
  W.~Luesomboon, N.~Ngo-Giang-Huong, K.~Chettra, T.~Cheunyam, T.~Suwarnlerk,
  S.~Ubolyam, W.~T. Shearer, R.~Paul, L.~M. Mofenson, L.~Fox, M.~G. Law, D.~A.
  Cooper, P.~Phanuphak, M.~C. Vun, and K.~Ruxrungtham.
\newblock Early versus deferred antiretroviral therapy for children older than
  1 year infected with {HIV} ({PREDICT}): a multicentre, randomised, open-label
  trial.
\newblock {\em Lancet Infectious Diseases}, 12(12):933--41, 2012.

\bibitem{Eglestion:2007}
B.~L. Egleston, D.~O. Scharfstein, E.~E. Freeman, and S.~K. West.
\newblock Causal inference for non-mortality outcomes in the presence of death.
\newblock {\em Biostatistics}, 8(3):526--545, 2007.

\bibitem{Chiba:2011}
Y.~Chiba and T.~J. VanderWeele.
\newblock A simple method for principal strata effects when the outcome has
  been truncated due to death.
\newblock {\em American Journal of Epidemiology}, 173(7):745--751, 2011.

\bibitem{Schwab:2016}
Joshua Schwab, Samuel Lendle, Maya Petersen, and Mark {van der Laan}.
\newblock {\em ltmle: Longitudinal Targeted Maximum Likelihood Estimation},
  2016.
\newblock {R} package version 0.9-9.

\bibitem{McCullagh:1989}
P.~McCullagh and J.~Nelder.
\newblock {\em Generalized Linear Models}.
\newblock Chapman and Hall/CRC, 1989.

\bibitem{Gruber:2010}
S.~Gruber and M.~J. van~der Laan.
\newblock A targeted maximum likelihood estimator of a causal effect on a
  bounded continuous outcome.
\newblock {\em International Journal of Biostatistics}, 6(1):Article 26, 2010.

\bibitem{Robins:2007}
James Robins, Mariela Sued, Quanhong Lei-Gomez, and Mark~Andrea Rotnitzky.
\newblock Comment: Performance of double-robust estimators when inverse
  probability weights are highly variable.
\newblock {\em Statistical Science}, 22(4):544--559, 2007.

\bibitem{Wood:2006}
S.~N. Wood.
\newblock {\em Generalized additive models: an introduction with R}.
\newblock Chapman and Hall/CRC, 2006.

\bibitem{Breimann:1985}
L.~Breiman and J.~H. Friedman.
\newblock Estimating optimal transformations for multiple-regression and
  correlation.
\newblock {\em Journal of the American Statistical Association},
  80(391):580--598, 1985.

\bibitem{Westreich:2012}
D.~Westreich, S.~R. Cole, J.~G. Young, F.~Palella, P.~C. Tien, L.~Kingsley,
  S.~J. Gange, and M.~A. Hernan.
\newblock The parametric g-formula to estimate the effect of highly active
  antiretroviral therapy on incident {AIDS} or death.
\newblock {\em Statistics in Medicine}, 31(18):2000--2009, 2012.

\bibitem{Petersen:2012}
M.~L. Petersen, K.~E. Porter, S.~Gruber, Y.~Wang, and M.~J. van~der Laan.
\newblock Diagnosing and responding to violations in the positivity assumption.
\newblock {\em Statistical Methods in Medical Research}, 21(1):31--54, 2012.

\bibitem{Sofrygin:2016}
Oleg Sofrygin, Mark~J. {van der Laan}, and Romain Neugebauer.
\newblock {\em simcausal: Simulating Longitudinal Data with Causal Inference
  Applications}, 2016.
\newblock {R} package version 0.5.3.

\bibitem{Schomaker:2019}
M.~Schomaker and C.~Heumann.
\newblock When and when not to use optimal model averaging.
\newblock {\em Statistical Papers}, in press, 2019.

\end{thebibliography}
}

\end{document}